\documentclass[preprint,aps,showpacs,nofootinbib,preprintnumbers,amsmath,amssymb]{revtex4-1}
\usepackage{amssymb}
\usepackage{epsfig}
\usepackage{graphicx}
\usepackage{subfigure}
\usepackage{dcolumn}
\usepackage{bm}
\usepackage[usenames ,dvipsnames]{xcolor}
\usepackage{slashed}

\begin{document}

\title{Direct Detection of Exothermic Dark Matter with  Light Mediator}

\author{Chao-Qiang~Geng$^{1,2,3}$\footnote{geng@phys.nthu.edu.tw}, 
Da~Huang$^{2}$\footnote{dahuang@phys.nthu.edu.tw}, Chun-Hao~Lee$^{2}$\footnote{lee.chunhao9112@gmail.com}, and Qing Wang$^{4,5}$\footnote{wangq@mail.tsinghua.edu.cn}
}
 \affiliation{$^{1}$Chongqing University of Posts \& Telecommunications, Chongqing, 400065, China\\
  $^{2}$Department of Physics, National Tsing Hua University, Hsinchu, Taiwan\\
  $^{3}$Physics Division, National Center for Theoretical Sciences, Hsinchu, Taiwan \\
  $^{4}$Department of Physics, Tsinghua University, Beijing, 100084, China\\
  $^{5}$Collaborative Innovation Center of Quantum Matter, Beijing, 100084, China }

\date{\today}

\begin{abstract}
We study the dark matter (DM) direct detection for the models with the effects of the isospin-violating couplings, 
exothermic scatterings, and/or the lightness of the mediator, proposed to relax the tension between the CDMS-Si signals 
and null experiments. In the light of the new updates of the LUX and CDMSlite data, we find that many of the previous 
proposals are now ruled out, including the Ge-phobic exothermic DM model and the Xe-phobic DM one with a light mediator. 
We also examine the exothermic DM models with a light mediator but without the isospin violation, and we are unable to 
identify any available parameter space that could simultaneously satisfy all the experiments. The only models that can 
partially relax the inconsistencies are the Xe-phobic exothermic DM models with or without a light mediator. 
But even in this case, a large portion of the CDMS-Si regions of interest has been constrained by the LUX and SuperCDMS data.
\end{abstract}

\maketitle

\section{Introduction}
\label{s1}
Although the existence of dark matter (DM) has been firmly established by many astrophysical 
and cosmological observations~\cite{PDG, Planck}, its particle nature remains a great mystery 
in modern particle physics. The leading DM candidate,  the so-called weakly interacting massive 
particle (WIMP)~\cite{Steigman:1984ac,Jungman:1995df,Feng:2010gw}, may be directly detected 
by its interactions with ordinary matter~\cite{Goodman:1984dc,FormFactor}. 
The possible signal for such interactions would be some nuclear recoils with energies 
of ${\cal O}({\rm keV})$ deposited in the detector. 

In the recent years, several DM direct detection experiments have reported potential signals for 
light WIMPs with masses around $1 \sim 10$ GeV, including DAMA~\cite{Bernabei:2008yi,Bernabei:2010mq}, 
CoGeNT~\cite{Aalseth:2010vx,Aalseth:2011wp,Aalseth:2012if,Aalseth:2014eft,Aalseth:2014jpa},  
CRESST-II~\cite{CRESST-II-S} and CDMS-Si~\cite{CDMS-Si}, while other experiments, 
such as  LUX~\cite{LUX2013, LUX2015}, SuperCDMS~\cite{SuperCDMS_Ge}, CDMSlite~\cite{Agnese:2013jaa, Agnese:2015nto}, 
XENON10~\cite{XENON10}, XENON100~\cite{XENON100, XENON100a}, CDEX~\cite{CDEX14, CDEX16} 
and PandaX~\cite{PandaX14, PandaX16}, have only presented null results. 
In order to reconcile the conflicts among the experiments, several mechanisms have been discussed, 
among which the isospin-violating interactions~\cite{Kurylov:2003ra,Giuliani:2005my,Feng:2011vu,Cirigliano:2013zta, Savage:2008er,IVDM1,Frandsen:2011cg, Cline:2011zr, Gao:2011ka, Chen:2014noa}, 
exothermic scatterings~\cite{Batell:2009vb,Graham:2010ca,Fox:2013pia,McCullough:2013jma,Frandsen:2014ima,Chen:2014tka,Gelmini:2014psa} 
and the light WIMP-nuclus mediators~\cite{Li:2014vza,Yang:2016wrl} are the three main proposals. 
In particular, after the releases of the 2013 LUX~\cite{LUX2013} and  2014 SuperCDMS~\cite{SuperCDMS_Ge} data, 
it was found~\cite{DelNobile:2013gba,Gelmini:2014psa,Cirigliano:2013zta,Gresham:2013mua, Chen:2014tka} 
that any of the single aforementioned mechanisms could not account for the inconsistencies in the DM direct searches, 
whereas only the combinations of two mechanisms above could totally or partially relax the tensions, 
such as the exothermic DM with the Xe-phobic~\cite{Chen:2014tka} or Ge-phobic~\cite{Gelmini:2014psa} 
interactions, and the Xe-phobic DM with a light mediator~\cite{Li:2014vza}. 
Recently, since the LUX~\cite{LUX2015} and CDMSlite~\cite{Agnese:2015nto} collaborations 
have updated their new measurements of the WIMP-nucleus recoil spectra, 
it is necessary to investigate the viability of the previous proposals in the light of new data. 
It is also interesting to explore some new schemes, such as the exothermic DM with nuclear scatterings mediated 
by a light particle with or without isospin-violating couplings, which has not been considered so far in the literature. 
Note that 
there are many proposals based on the general effective operators~\cite{Chang:2009yt, Fan:2010gt, Fitzpatrick:2012ix, 
Fitzpatrick:2012ib, Anand:2013yka, Gresham:2013mua, Catena:2016hoj}, 
aimed to comprehensively assess the compatibility of the positive signals and exclusion limits given by different experiments. 
Furthermore, it has been pointed out in Ref.~\cite{McCullough:2013jma} that the double-disk DM model~\cite{DDDM1,DDDM2} has the potential to improve the situation through the variation of the DM velocity distribution in our Galaxy.  
In this paper, we concentrate on DM models with two-component WIMPs
interacting with the detector nuclei via a dark photon, which could give a natural framework to incorporate the three mechanisms: 
the isospin-violating couplings, exothermic scatterings and a light mediator.  
We note that our specific model is chosen to illustrate the possible tendency for the DM direct search, while the similar low-energy behaviors 
could be extended to some general DM models.

The paper is organized as follows. In Sec.~\ref{Sec_Model}, 
we present 
the general framework for our DM models.
In Sec.~\ref{Sec_Fit} and Appendix~\ref{OtherExp}, we show our fitting methods 
for CDMS-Si, DAMA, and CoGeNT, and the procedures to obtain the exclusion limits for several relevant null experiments. 
In Sec.~\ref{Sec_Res}, we display our numerical results for our WIMP models with the several combined mechanisms. 
Finally, we summarize our main results in Sec.~\ref{Sec_Conc}. 

\section{General Framework for Dark Matter Direct Detection}\label{Sec_Model}
In the DM direct detection, one tries to measure the recoil energy deposited by the interaction of a WIMP particle with a nucleus in the detector~\cite{Goodman:1984dc}. As mentioned in Introduction, there have been several mechanisms invented to reconcile the tension between the positive signal in CDMS-Si and other null experiments, including the isospin-violation couplings~\cite{Feng:2011vu}, DM down-scatterings~\cite{Graham:2010ca,Frandsen:2014ima,Fox:2013pia,McCullough:2013jma}, and the introduction of a light mediator~\cite{Li:2014vza}. In the present paper, we would like to provide a unified framework to incorporate all these three effects, which is the variation of the isospin-violating DM model with a light dark photon proposed in Refs.~\cite{Frandsen:2011cg, Cline:2011zr, Gao:2011ka, Chen:2014noa,Frandsen:2014ima, Chen:2014tka}. 

\subsection{The Benchmark Dark Matter Model}
For concreteness, we assume that 
DM in our Universe is composed of two Majorana fermionic WIMP particles, $\chi_H$ and $\chi_L$, 
with a small mass difference denoted by $\delta \ll m_{H/L}$. The scattering of a WIMP particle 
to a nucleus in the detector is mediated by a light dark photon $\phi$ with its mass $m_\phi$. 
Due to their Majorana nature, the two WIMP particles can only couple to the dark photon off-diagonally in the mass basis via the interaction:
\begin{eqnarray}
- f_\chi^V (\bar{\chi}_H \slashed{\phi} \chi_L + \bar{\chi}_L \slashed{\phi} \chi_H)\,,
\end{eqnarray}
resulting that the elastic scatterings between WIMP particles and nucleons appear at one-loop order 
and the inelastic scatterings would dominate the WIMP-nucleon interactions~\cite{Chen:2014tka}. 
We also assume that the dark photon couples to the SM quarks via the vector current $-f_q^V \bar{q} \slashed{\phi} q$. 
Note that this choice of the couplings would only lead to the spin-independent DM direct detection signals. 

Generically, one expects that the two WIMP particles both carry half of the DM relic density in our Universe, 
so that we have two scattering processes: the up-scattering from the lower-energy state $\chi_L$ to 
its higher-energy partner $\chi_H$, and the down-scattering from $\chi_H$ to $\chi_L$. 
However, it has been pointed out in Ref.~\cite{Graham:2010ca} that only the high-velocity tail of 
the dark matter distribution in our galaxy has the enough energy to up-scatter. As a result,  
 the down-scattering dominates the WIMP-nucleon event rates. In the following discussions, 
 we will concentrate on the WIMP down-scattering off a target nucleus $T$ as
\begin{eqnarray}
\chi_H(p_1) + T(p_2) \to \chi_L(p_3) + T(p_4)\,.
\end{eqnarray} 
Due to the rest mass change of the WIMP particles, the required minimum velocity 
to produce a nuclear recoil of the energy $E_{\rm nr}$ is
\begin{eqnarray}\label{vMin}
v_{\rm min} = \frac{1}{\sqrt{2E_{\rm nr} m_T}} \left|\delta + \frac{m_T E_{\rm nr}}{\mu_{\chi T}} \right|
\end{eqnarray} 
where $m_T$ denotes the mass of a target nucleus, $\mu_{\chi T} = m_\chi m_T/(m_\chi+m_T)$ 
represents the reduced mass of the WIMP-nucleus system, and $\delta = m_L- m_H < 0$ is 
for exothermic DM scatterings following the convention in the literature~\cite{Graham:2010ca,Frandsen:2014ima,Fox:2013pia}. 

Note that when the mediator mass $m_\phi$ is much larger than the 3-momentum transfer 
$q = |{\bf q}| = |{\bf p}_3 - {\bf p}_1|$ in the WIMP-nucleus scattering, the interaction between the WIMP 
and a nucleon $N=(p,n)$ can be effectively described by the local operator, given by
\begin{eqnarray}\label{EOp}
{\cal O}_0 = \frac{c_N}{m_\phi^2} (\bar{\chi}_H\gamma^\mu \chi_L+\bar{\chi}_L \gamma^\mu \chi_H) (\bar{N}\gamma_\mu N)\,,
\end{eqnarray}
where the Wilsonian coefficient $c_N$ can be derived from more fundamental parameters $f_\chi^V$ and $f_q^V$. 
However, such a description is no longer valid when the value of $m_\phi$ is comparable to 
or even smaller than the typical momentum transfer $q$. As discussed in Refs.~\cite{Li:2014vza,Yang:2016wrl}, 
the appropriate effective operator  should be formulated as
\begin{eqnarray}\label{EOm}
{\cal O} = \frac{c_N}{q^2 + m_\phi^2} (\bar{\chi}_H\gamma^\mu \chi_L+\bar{\chi}_L \gamma^\mu \chi_H) (\bar{N}\gamma_\mu N)\,,
\end{eqnarray}
in which the lightness of the dark photon is accounted for by its complete propagator. 
Consequently, the differential cross section of the WIMP-nucleon cross section can be written as~\cite{Li:2014vza}
\begin{eqnarray}
\frac{d \sigma_N}{dq^2} (q^2, v) = \frac{\bar{\sigma}_N}{4 \mu_{\chi N}^2 v^2} G(q^2, v)\,,
\end{eqnarray}
where $\mu_{\chi N}$ is the nucleon-WIMP reduced mass,  $\bar{\sigma}_N$ is the reference cross section 
defined at the typical velocity $v_{\rm ref} = 200~{\rm km \cdot s}^{-1}$, 
and $G(q^2, v)$ at the cross section level is defined by
\begin{eqnarray}\label{G_def}
G(q^2, v) = \frac{(q^2_{\rm ref} - q^2_{\rm min})\overline{|{\cal M}_{\chi N}(q^2, v)|^2}}{\int^{q_{\rm ref}^2}_{q_{\rm min}^2} d q^2 \overline{|{\cal M}_{\chi N}(q^2, v_{\rm ref})|^2} } \,,
\end{eqnarray}
which reflects the deviation from the standard point-like interaction of Eq.~(\ref{EOp}) 
due to the extra momentum transfer dependence in Eq.~(\ref{EOm}). In Eq.~(\ref{G_def}), 
$\overline{|{\cal M}_{\chi N}|^2}$ is the squared matrix element averaged over the initial state spins, 
$q^2_{\rm ref} = 4\mu_{\chi N}^2 v_{\rm ref}^2$, and $q^2_{\rm min}$ is related to the energy thresholds of 
DM direct detection experiments with typical values of ${\cal O}({\rm keV}^2)$.  
Note that our effective operator in Eq.~(\ref{EOm}) corresponds to the vector-like interaction ${\cal O}_5$ 
defined in Ref.~\cite{Li:2014vza}, so that  $G(q^2, v)$ can be reduced to the following simple form:
\begin{eqnarray}\label{G1Factor}
G (q^2) = \frac{(1+q_{\rm min}^2/m_\phi^2)(1+ q_{\rm ref}^2/m_\phi^2)}{(1+ q^2/m_\phi^2)^2}\,. 
\end{eqnarray}
Usually, $q_{\rm min}^2$ is too small compared to other scales in the formula, 
so we effectively choose it to be zero for simplicity. It is also evident that, 
when the mediator becomes heavy, i.e., $m_\phi^2 \gg q^2$ and $q_{\rm ref}^2$, 
the factor $G$ approaches unity, so that we recover the conventional WIMP-nucleon cross section for contact interactions.
 
With the modifications above, the spin-independent WIMP-nucleus differential cross section can be expressed as
\begin{eqnarray}
\frac{d\sigma_T}{d q^2} = \frac{m_T}{2\mu_{\chi p}^2 v^2} \bar{\sigma}_p [Z+\xi(A-Z)]^2 G (q^2) F^2_T (q^2)\,,
\end{eqnarray}  
where $A$ ($Z$) is the mass (atomic) number of the target, 
$\xi$ is the coupling ratio between the neutron and proton, 
and $F_T(q^2)$ is the nuclear form factor taken as the conventional Helmi form~\cite{FormFactor}: 
\begin{eqnarray}
F(q) = 3 e^{-q^2 s^2 /2} \frac{\sin(qr)-qr\cos(qr)}{(qr)^3}\,,
\end{eqnarray}
where $s = 0.9$~fm and $r$ is the effective nuclear radius, given by~\cite{FormFactor},
\begin{eqnarray}
r = \sqrt{c^2 + \frac{7}{3}\pi^2 a^2 -5s^2}\,,
\end{eqnarray} 
with $c = 1.23 A^{1/3}-0.60$~fm and $a = 0.52$~fm. 
If $\xi = 1$, then the WIMP particles couple to the nucleon universally as the conventional isospin-conserving DM. 
However, when $\xi \neq 1$, the interaction strengths of WIMPs with a proton and a neutron are distinctive, 
realizing the so-call isospin-violation 
DM~\cite{Kurylov:2003ra,Giuliani:2005my,Feng:2011vu,Cirigliano:2013zta, Savage:2008er, IVDM1,Frandsen:2011cg}. 
In particular, when $\xi = -0.7$, the constraints from direct search experiments with the liquid xenon target are weakened maximally, while when $\xi = -0.8$, the germanium detector would lose its sensitivity mostly. In the literature, the above two classes of models are usually called the ``Xe-phobic"~\cite{Chen:2014tka} and ``Ge-phobic''~\cite{Gelmini:2014psa} DM, respectively. It is shown in Refs.~\cite{Frandsen:2011cg, Cline:2011zr, Gao:2011ka} that such isospin-violating WIMP interactions can be achieved via the kinetic and mass mixings between the dark photon and Standard Model $U(1)$ gauge fields, as well as direct couplings of the dark photon with SM quarks.

Note that we only focus on the direct detections for the benchmark WIMP model with a dark photon in the present paper. For other aspects of the model, we expect that the phenomenologies are similar to those discussed in Refs.~\cite{Frandsen:2011cg, Cline:2011zr, Gao:2011ka}, especially for the generation of the DM relic abundance in the Universe. Furthermore, we remark that our direct detection results can actually be applied to other scalar or fermionic WIMP models corresponding to the generalized type-I effective operators defined in Ref.~\cite{Li:2014vza}, since they would give rise to the same factor $G(q^2,v)$ in Eq.~(\ref{G1Factor}) in the non-relativistic limit.  

\subsection{Recoil Rates in Dark Matter Direct Searches }
The differential recoil event rate per unit detector mass for only one isotope $T$ is given by:
\begin{eqnarray}
\frac{dR}{dE_{\rm nr}} &=& \frac{dN}{M_T dt dE_{\rm nr}} = \frac{\rho_\chi}{m_\chi} \int_{|\mathbf{v}|>v_{\rm min}} d^3 \mathbf{v} v f(\mathbf{v}) \frac{d \sigma_T}{dq^2} \nonumber\\
&=& \frac{\rho_\chi}{2 m_\chi \mu_{\chi p}^2} \bar{\sigma}_p [Z + (A-Z)\xi]^2 G(E_{\rm nr}) F_A^2(E_{\rm nr}) \eta(E_{\rm nr}, t) \,,
\end{eqnarray}
where $\rho_\chi = 0.3$~GeV/${\rm cm}^3$ is the local DM energy density, and we have transformed the dependence of the momentum transfer $q^2$ to the experimentally measured recoil energy $E_{\rm nr}$ via the relation $E_{\rm nr} = q^2/(2m_T)$. For the present model, the only WIMP velocity dependence in the differential cross section $d\sigma_T/dq^2$ is proportional to $1/v^2$, so that the information of the DM velocity distribution is included in the factor $\eta(E_{\rm nr},t)$ defined as
\begin{eqnarray}
\eta(E_{\rm nr}, t) = \int_{|{\mathbf v}|>v_{\rm min}} d^3 {\bf v} \frac{f({\mathbf v})}{v}\,,
\end{eqnarray}
which has the natural interpretation of the mean inverse speed of the WIMP particles in the local Earth frame, with the minimum velocity $v_{\rm min}$ for the integration taken as in Eq.~(\ref{vMin}) to account for the inelastic nature of scatterings.

For the DM velocity distribution $f_G(\mathbf{v})$ in our Galaxy, we follow the Standard Halo Model (SHM)~\cite{Freese:1987wu, Savage:2008er}, which is assumed to be a truncated Maxwell-Boltzmann distribution with a cutoff at $v_{\rm esc} = 544~{\rm km/s}$~\cite{Smith:2006ym},
given by
\begin{eqnarray}
f_G({\mathbf v}) = \frac{1}{N_{\rm esc} (\pi v_0^2)^{3/2}} e^{-{\mathbf v}^2/v_0^2} \Theta (v_{\rm esc} -v)\,,
\end{eqnarray}
where $v_0=220~{\rm km/s}$ is the mean DM velocity relative to the Galaxy, $v_{\rm esc}$ 
is the Galactic escape velocity, and $N_{\rm esc}$ is the normalization factor, given by
\begin{eqnarray}
N_{\rm esc} = {\rm erf}(z) -2 z \exp(-z^2)/\pi^{1/2}\,, 
\end{eqnarray}    
with $z \equiv v_{\rm esc}/v_0$. By taking into account the motion of the Earth relative to the Galactic halo rest frame with the speed of ${\bf v}_e$, the observed WIMP velocity distribution should be obtained through a Galilean transformation
\begin{eqnarray}\label{DMVelocity}
f({\mathbf v}) = f_G({\bf v} + {\bf v}_e)\,.
\end{eqnarray}
Due to the motion of the Earth around the Sun, we can write ${\mathbf v}_e$ as
\begin{eqnarray}
v_e = v_{\odot} + v_{\rm orb} \cos\gamma \cos[\omega(t-t_0)]\,,
\end{eqnarray}
where $\omega = 2\pi/$year, $\cos \gamma = 0.51$, $v_\odot = 232~{\rm km/s}$ is the Sun's motion with respect to the WIMP component rest frame, and $v_{\rm orb} = 29.8~{\rm km/s}$ is the Earth orbital speed. With this WIMP velocity distribution, the mean inverse speed $\eta(E_{\rm nr},t)$ can be analytically computed:
\begin{eqnarray}
\eta(E_{\rm nr}, t) = \left\{ \begin{array}{ll}
\frac{1}{v_0 y}\,,   & {\rm for}~z<y,~x<|y-z|   \\
\frac{1}{2N_{\rm esc} v_0 y} \left[ {\rm erf}(x+y) -{\rm erf}(x-y) - \frac{4}{\sqrt{\pi}}y e^{-z^2} \right]\,, & {\rm for}~z>y, ~x<|y-z| \\
\frac{1}{2N_{\rm esc} v_0 y} \left[ {\rm erf}(z) -{\rm erf}(x-y) - \frac{2}{\sqrt{\pi}}(y+z-x) e^{-z^2} \right]\,, & {\rm for} ~|y-z|<x<y+z\\
0, & {\rm for}~y+z<x\\
\end{array}
\right.\nonumber\\
\end{eqnarray}
where $x\equiv v_{\rm min}/v_0, \quad y \equiv v_e/v_0$, and $z \equiv v_{\rm esc}/v_0$.

For different DM direct detection experiments, the measured signals of $s$ can be vastly  different, such as the prompt scintillation signal $S1$, the ionization charge signal $S2$, the phonons released, the electron equivalent energy $E_{\rm ee}$, and so on. However, all these signals can be related to the WIMP nuclear recoil energy $E_{\rm nr}$ via some definite function $s = f_s(E_{\rm nr})$.  In general, the recoil rate per unit mass in terms of these observables can be written as~\cite{Li:2014vza,Savage:2008er}
\begin{eqnarray}
R(t) = \int^\infty_0 dE_{\rm nr} \epsilon(s) \Phi(f_s(E_{\rm nr}), s_1, s_2) \left(\frac{dR}{dE_{\rm nr}} \right) \,,
\end{eqnarray} 
where $\epsilon(s)$ denotes the efficiency of detecting a signal $s$, and $\Phi(f_s(E_{\rm nr}),s_1, s_2)$ is the response function corresponding to the probability of observing a signal $ f_s(E_{\rm nr})$ in the range $[s_1, s_2]$ given an energy $E_{\rm nr}$. By assuming that the measured value of the signal is normally distributed around $f_s(E_{\rm nr})$ with the standard deviation $\sigma(s)$ which is the resolution given by the experiment, we can obtain the analytical expression for $\Phi$ as follows:
\begin{eqnarray}\label{resolution}
\Phi (f_s(E_{\rm nr}),s_1, s_2) = \frac{1}{2}\left[{\rm erf}\left(\frac{s_2 -f_s(E_{\rm nr}) }{\sqrt{2}\sigma}\right) -\left( \frac{s_1 -f_s(E_{\rm nr}) }{\sqrt{2}\sigma}\right) \right]\,.
\end{eqnarray}

When the target material of a detector is composed of multiple elements or isotopes, the total event rate is given by
\begin{eqnarray}
R_{\rm tot} = \sum_i f_i R_i(t)\,,
\end{eqnarray} 
where $R_i$ denotes the rate for the element/isotope $i$ and $f_i$ is its mass fraction. In our discussion, we use the isotope number abundances for various elements listed in Table II of Ref.~\cite{Feng:2011vu} to calculate the isotope mass fractions in a detector. Finally, by multiplying the exposure Ex given by different experiments, we can obtain the expected number of recoils
\begin{eqnarray}
N_{\rm rec}  = {\rm Ex} \cdot R_{\rm tot} (t)\,.
\end{eqnarray}

\subsection{Annual Modulation}
Due to the motion of the Earth around the Sun, the nuclear recoils of WIMPs in the detector would experience an annual modulation~\cite{Drukier:1986tm, Freese:1987wu}, which is one of the key signals in the DM direct detection. For the SHM considered here, such modulations can be approximated to be~\cite{Savage:2008er}:
\begin{eqnarray}
S_m (E_{\rm nr}) = \frac{1}{2} \left[ \frac{dR}{dE_{\rm nr}} (E, ~{\rm June~1}) - \frac{dR}{dE_{\rm nr}} (E_{\rm nr}, ~{\rm Dec.~1}) \right]\,.
\end{eqnarray}
Usually, experiments such as DAMA~\cite{Bernabei:2008yi,Bernabei:2010mq} and CoGeNT~\cite{Aalseth:2014eft} often give the data for the average amplitude over some range $[E_1, E_2]$, given by
\begin{eqnarray}\label{Modulation}
S_m = \frac{1}{E_2-E_1} \int^{E_2}_{E_1} dE S_m(E_{\rm nr})\,.
\end{eqnarray}


\section{Signals and Constraints}\label{Sec_Fit}
In the recent decades, there have been numerous DM direct detection experiments to search for WIMP signals. 
Some of them have reported positive signals for light WIMPs 
such as DAMA~\cite{Bernabei:2008yi,Bernabei:2010mq}, CoGeNT~\cite{Aalseth:2010vx,
Aalseth:2011wp,Aalseth:2012if,Aalseth:2014eft,Aalseth:2014jpa},  CRESST-II~\cite{CRESST-II-S}, and CDMS-Si~\cite{CDMS-Si}, 
while others have only presented the exclusion limits on the WIMP-nucleon cross section, 
including LUX~\cite{LUX2013, LUX2015}, SuperCDMS~\cite{SuperCDMS_Ge}, 
CDMSlite~\cite{Agnese:2013jaa, Agnese:2015nto}, XENON10~\cite{XENON10}, 
XENON100~\cite{XENON100, XENON100a}, CDEX~\cite{CDEX14, CDEX16}, 
PandaX~\cite{PandaX14, PandaX16}, and so on. 
Presently, the most stringent constraints come from 
the data of SuperCDMS~\cite{SuperCDMS_Ge}, CDMSlite~\cite{Agnese:2015nto}, 
and LUX~\cite{LUX2013, LUX2015} for low-mass WIMPs, which strongly conflict
 with the signal regions derived from DAMA and CoGeNT. Moreover, the CRESST-II positive result 
 in Ref.~\cite{CRESST-II-S} has not been confirmed by the more recent data~\cite{CRESST-II} of the same CaWO$_4$ detector. 
 In addition, as pointed out in Refs.~\cite{Gelmini:2014psa,Chen:2014tka,Li:2014vza}, 
 only the CDMS-Si dataset has been found to be marginally consistent with the LUX 2013~\cite{LUX2013} 
 and SuperCDMS~\cite{SuperCDMS_Ge} ones with the Xe-phobic or Ge-phobic interactions. 
 With the recent updates by LUX~\cite{LUX2015} and CDMSlite~\cite{Agnese:2015nto},
  it is necessary to look into if this conclusion is still valid or not. 
  Therefore,  we focus on the compatibility of the CDMS-Si signal region and constraints 
  from LUX and CDMS experiments in the light of these new datasets, with the analysis details collected as follows:

\subsection*{CDMS-Si}
The CDMS-Si experiment~\cite{CDMS-Si} has measured both ionization electrons and phonons 
in its silicon detector with the raw exposure of 140.2~kg-days, and observed three candidate events 
of recoil energies at $E_{\rm nr}=8.2,~9.5$, and 12.3 keV, respectively, in the benchmark energy 
range $7\sim 100$~keV. Following Ref.~\cite{Cirigliano:2013zta}, we bin the data in 2 keV intervals, 
so that the three candidates lie in the first three bins. 
We adopt the efficiency for the data selection shown as the solid blue curve in Ref.~\cite{CDMS-Si}, 
and assume the resolution to be perfect. From the rescaled background distributions from Page 10 of Ref.~\cite{CDMS_Si_bkg}, 
we get that the backgrounds from surface events, $^{246}$Pb and neutrons are 0.41, 0.13 and 0.08 events, respectively. 
In order to obtain the best-fit regions, we maximize the log of the extended likelihood function~\cite{Barlow:1990vc} constructed as
\begin{eqnarray}
{\cal L} = e^{-(N+B)} \prod^n_{i} \left[\left(\frac{dN}{dE_{\rm nr}} \right)_i + \left(\frac{dB}{dE_{\rm nr}} \right)_i  \right]\,,
\end{eqnarray}
where $B(N)$ is the total expected number of signal (background) events in the whole recoil energy range, and $(dN(B)/dE_{\rm nr})_i$ the corresponding differential event rate at the $i$-th bin. The allowed parameter space regions at $68\%$ and $90\%$ C.L. are obtained with the contours satisfying $2 \Delta\ln {\cal L} = 2.3$ and 4.6, respectively. 

\subsection*{SuperCDMS}
SuperCDMS experiments~\cite{SuperCDMS_Ge} measure the ionization and phonon signals simultaneously, and they can effectively reject the background electric recoil events and enhance the sensitivity to the low-mass WIMP searches. We consider the low-energy data~\cite{SuperCDMS_Ge} in the range of $1.6 \sim 10$ keV$_{\rm nr}$ with the exposure of 577 kg-days. The cumulative efficiency is taken as the red curve in Fig.~1 of Ref.~\cite{SuperCDMS_Ge}, which is the efficiency after various data-selection criteria. We obtain the 90$\%$ C.L. exclusion limits on the spin-independent WIMP-nucleon cross section based on the $p_{\rm Max}$ method~\cite{Yellin:2002xd} with a perfect energy resolution and no background subtraction, which, from our perspective, is very similar to the optimum interval method adopted by the SuperCDMS Collaboration.

\subsection*{CDMSlite}
We use the CDMSlite data shown in Fig.~3 of Ref.~\cite{Agnese:2015nto} to obtain the 
constraint for the spin-independent cross sections. The experiment is very sensitive to the low-mass region 
due to its low threshold. The nuclear-recoil (NR) energy $E_{\rm nr}$ is related to the measured 
electron-equivalent energy $E_{\rm ee}$ via the transformation: 
\begin{eqnarray}
E_{\rm nr} &=& E_{\rm ee} \left( \frac{1+e V_b/\epsilon_\gamma}{1+ Y(E_{\rm nr})eV_b/\epsilon_\gamma} \right)\,,
\end{eqnarray}
where the voltage bias $V_b = -70$~V, and the average energy to produce a electron-hole pair in germanium 
is $\epsilon_\gamma = 3$~eV/pair. Following the Lindhard model~\cite{Lindhard, Sorensen:2011bd}, the ionization yield for NRs is given by
\begin{equation}
Y(E_{nr}) = \frac{k g(\epsilon)}{1+ k g(\epsilon)}\,,
\end{equation} 
where $k = 0.157$, $g(\epsilon) = 3 \epsilon^{0.15}+0.7 \epsilon^{0.6}+\epsilon$, 
$\epsilon = 11.5 E_{\rm nr}({\rm keV})Z^{-7/3}$ and $Z_{\rm Ge} = 32$ is the atomic number of germanium nuclei. 
The total analysis exposure is 70.1 kg-days, with the live time for the same SuperCDMS iZIP detector 97.81 days of 
period I and 17.78 days of period II. The thresholds are different for two periods, which are 75 and 56 ${\rm keV_{ee}}$, 
respectively. The total signal efficiency can be read out from Fig.~1 of Ref.~\cite{Agnese:2015nto}. The resolution of the detector is obtained by extrapolating the resolutions 0.10, 0.03, and 0.018 keV$_{ee}$ of the three $^{71}{\rm Ge}$ electron-capture peaks at 10.37, 1.30, and 0.16 keV$_{ee}$, respectively, which are given in Table I of Ref.~\cite{Agnese:2015nto}. In the present paper, we exploit the $p_{\rm Max}$ method~\cite{Yellin:2002xd} with no background subtraction to obtain $90\%$ C.L. upper limits for spin-independent WIMP-nucleon cross sections with different WIMP masses.

\subsection*{LUX2013}
The LUX experiment uses a dual-phase xenon time-projection chamber which measures 
both the primary scintillation light $S1$ and ionization charge $S2$. The exposure of 
the LUX2013 data in Ref.~\cite{LUX2013} is $118$ kg $\times$ 85.3 live days. Following Ref.~\cite{PLRtest}, 
the event number in the signal range $S1 \in [S1_a, S1_b]$ is given by:
\begin{eqnarray}\label{NLUX}
N_{[S1_a, S1_b]} &=& {\rm Ex} \int^{S1_b}_{S1_a} dS1 \Big[ \sum^{\infty}_{n=1} \epsilon(S1) {\rm Gauss}(S1|n,\sqrt{n}\sigma_{PMT}) \nonumber\\
& & \times \int^\infty_0 {\rm Poiss}(n|\nu(E_{\rm nr}))\epsilon_{S2}(E_{\rm nr})\frac{dR}{dE_{\rm nr}} dE_{\rm nr} \Big]\, ,
\end{eqnarray}
with the average expected number of photoelectrons $\nu(E_{\rm nr})$ for the nuclear recoil energy $E_{\rm nr}$ to be 
\begin{equation}\label{S1transf}
\nu(E_{\rm nr}) = E_{\rm nr} {\cal L}_{\rm eff} (E_{\rm nr}) L_y \frac{S_{\rm nr}}{S_{ee}}\, ,
\end{equation}
where ${\cal L}_{\rm eff}$ is the energy-dependent scintillation efficiency of the liquid xenon, $L_y$ the light yield, 
and $S_{\rm nr}$ ($S_{\rm ee}$) the nuclear-(electron-)recoil quenching factor. 
We adopt the energy-dependent absolute light yield, ${\cal L}_{\rm eff} L_y S_{\rm nr}/S_{\rm ee}$, 
from the slide 25 in Ref.~\cite{LUXLightYield}, with a hard cutoff at 3 keV. 
For the DM detection efficiency $\epsilon(S1)$, we interpolate the black-upward-triangle curve 
in the lower panel of Fig.~1 in Ref.~\cite{LUX2013}, as used by the LUX Collaboration 
in its profile likelihood analysis. An additional $S2$ efficiency $\epsilon_{S2}(E_{\rm nr}) = \Theta (E_{\rm nr} - 3 {\rm keV})$ 
is taken into account. 

Note that only one event at $(S1, \log(S2_b/S1)) = (3.2, 1.75)$ marginally passes the mean of the Gaussian fit to the simulated WIMP nuclear recoil data in the $S1$-$\log(S2_b/S1)$ plane, which is the solid red curve of Fig.~4 in Ref.~\cite{LUX2013}. In our work, we set the 90$\%$ C.L. upper limits by performing the simple maximum gap analysis~\cite{Yellin:2002xd} for the signal region of 2-30 phe and taking this single candidate event into account. 

\subsection*{LUX2015}
Recently, the LUX Collaboration has updated the WIMP direct search in Ref.~\cite{LUX2015}. 
In this new data analysis, the LUX Collaboration has calibrated the photomultiplier tube (PMT) 
signals for both the scintillation photons and ionization electrons in units of detected photons (phd), 
rather than the photoelectrons used in the first analysis~\cite{LUX2013}. 
In addition, the new analysis made the advantages of the lowered thresholds, 
more stringent energy cutoffs, larger exposures, and better signal resolutions, 
so that even stronger spin-independent WIMP-nucleon cross sections could be placed in the low WIMP mass region. 
The exposure is 154.4~kg $\times$ 95.0 live-days, and the nuclear recoil energies can be calibrated down to 1.1 keV, 
with the highest endpoint at 18.6 keV.

In our work, we apply almost the same method as that for the LUX2013 dataset by approximating the event number in the signal range $S1 \in [S1_a, S1_b]$ in Eq.~(\ref{NLUX}). For the $S1$ signal in terms of units of phd, we obtain the number of photons $n_\gamma$ for the NR energy $E_{\rm nr}$ with the photon yield given in the middle plot of Fig.~1 in Ref.~\cite{LUX2015}, and then transform it into $S1$ with the relation $\langle S1 \rangle = g_1 n_\gamma$ through the LUX-specific gain factor $g_1 = 0.117$~phd. The resolution for the LUX experiments is $\sigma_{\rm PMT} = 0.2$~phd. For the detection efficiency, we adopt the one shown as the black solid curve in the bottom plot of Fig.~1 in Ref.~\cite{LUX2015}. We also take into account an additional efficiency of $97.5\%$ due to the pulse-identification. The benchmark signal region is chosen as that below the red solid line in Fig.~2 of Ref.~\cite{LUX2015}, with the assumption of $50\%$ efficiency~\cite{Catena:2016hoj,Gresham:2013mua}. By construction, this assumption is well-motivated for large DM masses, while it undercounts the signal events at the low mass region, so that the obtained upper limits are conservative. In our numerical calculation, we perform the Poisson statistics test to give the 90$\%$ C.L. upper bounds by assuming that there is not any candidate event in the signal region, due to the fact that the data after cuts agrees with the background-only model very well.

For comparison, we also take into account the DAMA~\cite{Bernabei:2010mq} and CoGeNT~\cite{Aalseth:2012if} signal regions as well as the WIMP exclusion limits from XENON10~\cite{XENON10}, XENON100~\cite{XENON100} and CDEX-1~\cite{CDEX14}, with the analysis details listed in Appendix~\ref{OtherExp}. Since these experiments are not our main focus, we only present their results for the reference WIMP model with the isospin-conserving elastic point-like nuclear scatterings, as well as some cases when the models  have the potential to reconcile the tension between CDMS-Si and null results from LUX, SuperCDMS and CDMSlite.

\section{Fitting Results}\label{Sec_Res}
\begin{figure}[ht]
\includegraphics[scale = 0.6, angle=270]{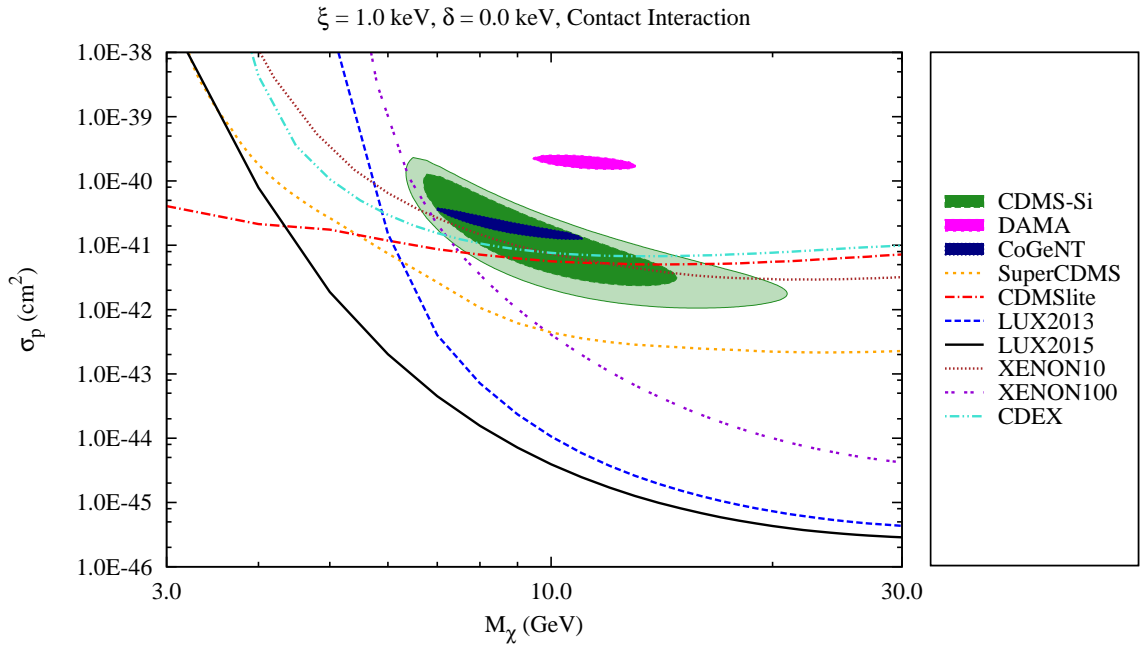}
\caption{Regions of interest for CDMS-Si at $68\%$ (dark green) and $90\%$ (light green) C.L., DAMA (magenta) and CoGeNT (dark blue) at $90\%$ C.L., and $90\%$ exclusion curves for SuperCDMS (yellow dotted), CDMSlite (red dot-dashed), LUX2013 (blue dashed), LUX2015 (black solid), XENON10 (brown densely dashed), XENON100(purple loosely dotted), and CDEX (cyan doubly-dotted-dashed), which are relevant to the standard spin-independent isospin-conserving elastic nucleon-WIMP scattering with a contact interaction. }\label{ResConv}
\end{figure}

In this section, we present  our results for different WIMP models, 
which combine the effects of the isospin-violating couplings, exothermic scatterings, 
and the lightness of the mediator. We first discuss the conventional spin-independent WIMP model with isospin-conserving elastic scatterings via a contact interaction. The result is shown in Fig.~\ref{ResConv}, which includes the fits to the DAMA~\cite{Bernabei:2010mq}, CoGeNT~\cite{Aalseth:2012if}, and CDMS-Si~\cite{CDMS-Si} datasets, as well as constraints from LUX~\cite{LUX2013, LUX2015}, SuperCDMS~\cite{SuperCDMS_Ge}, CDMSlite~\cite{Agnese:2015nto}, XENON10~\cite{XENON10}, and XENON100~\cite{XENON100}.
It is evident that the most stringent constraints for the CDMS-Si regions of interest come from SuperCDMS~\cite{SuperCDMS_Ge}, CDMSlite~\cite{Agnese:2015nto}, and LUX~\cite{LUX2013, LUX2015}. In particular, the recent improvement by LUX in the sensitivity of the low-mass WIMP region has made the tension with the CDMS-Si signals more severe. 

Note that after the releases of the LUX2013~\cite{LUX2013} and  SuperCDMS~\cite{SuperCDMS_Ge} data, 
it was found~\cite{DelNobile:2013gba,Gelmini:2014psa,Cirigliano:2013zta,Gresham:2013mua, Chen:2014tka} that any single mechanism from the isospin violation, exothermic interaction, and a light mediator could not reconcile the tension any more, so that several combinations were exploited to make the CDMS-Si signal region compatible with other null experiments, such as the Xe-phobic exothermic DM in Ref.~\cite{Chen:2014tka}, the Ge-phobic exothermic DM in Ref.~\cite{Gelmini:2014psa}, and the Xe-phobic elastic DM with a light mediator in Ref.~\cite{Li:2014vza}. However, as illustrated in Fig.~\ref{ResConv}, the recent results from LUX~\cite{LUX2015} and CDMSlite~\cite{Agnese:2015nto} improve the sensitivity to low-mass WIMPs significantly, which clearly put additional challenges to the existing reconciliation mechanisms. Therefore, it is timely and necessary to revisit the above DM models in the light of the new LUX and CDMSlite data. We also want to study the compatibility of the above experiments in the context of an isospin-conserving/isospin-violating exothermic WIMP model with the nucleon scattering induced by a light mediator, which has not been explored yet.

\subsection{Exothermic DM with Isospin Violation}
\begin{figure}[ht]
\includegraphics[scale = 0.32, angle=270]{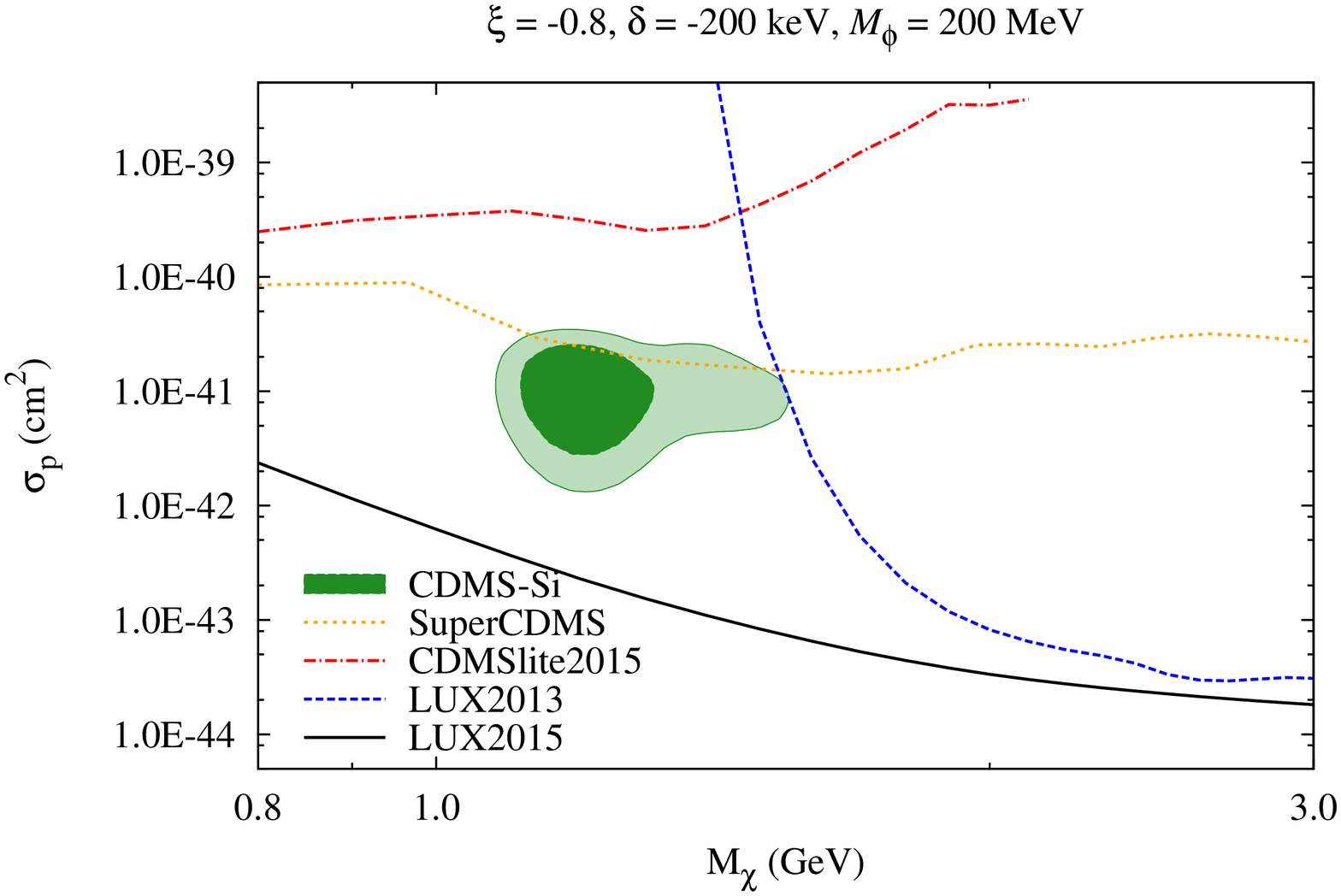}
\includegraphics[scale = 0.32, angle=270]{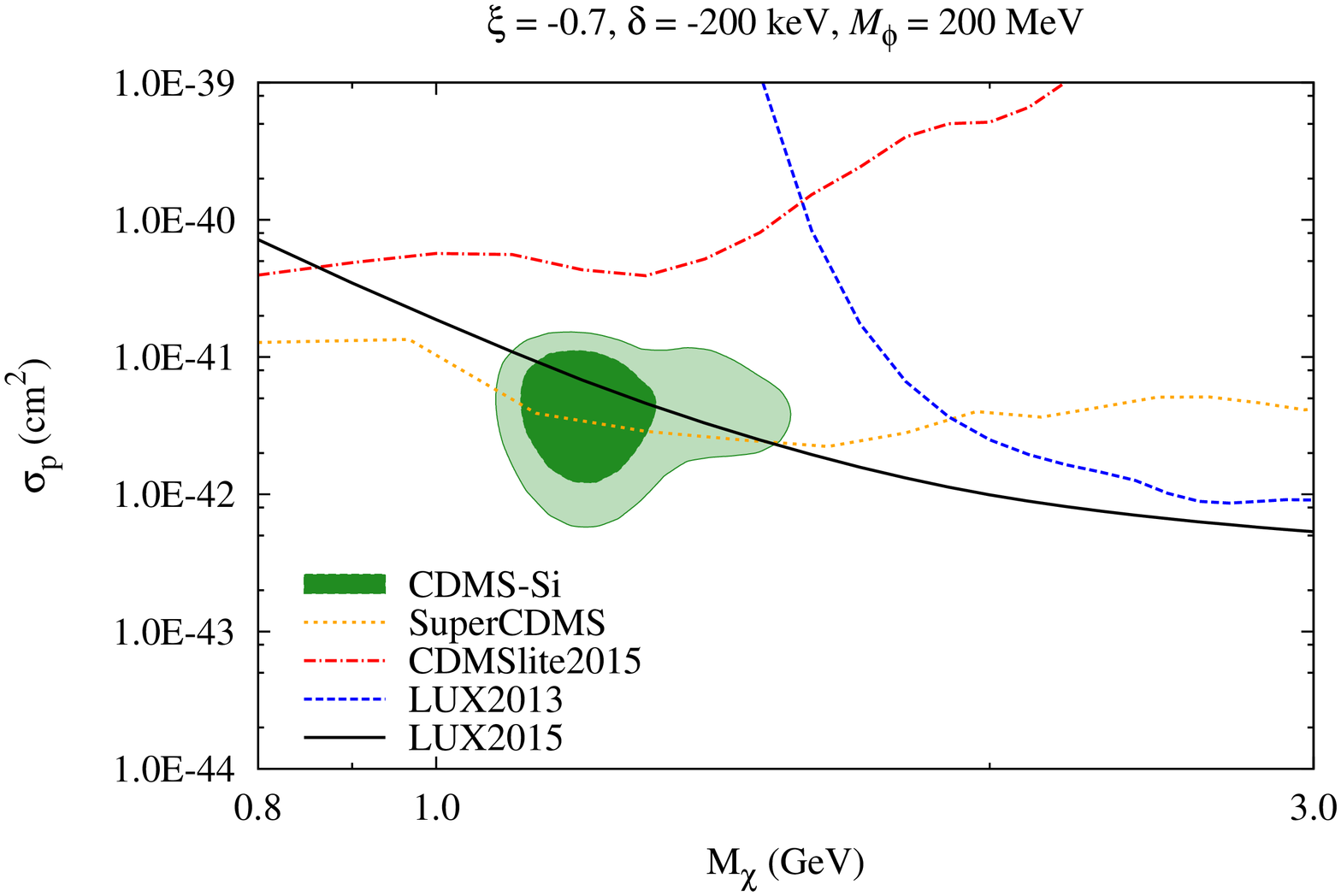}
\caption{The CDMS-Si $68\%$ (dark green) and $90\%$ (light green) C.L. regions of interest and $90\%$ C.L. exclusion curves for SuperCDMS (yellow dotted), CDMSlite (red dot-dashed), LUX2013 (blue dashed), and LUX2015 (black solid) experiments, where the left (right) graph corresponds to the Ge(Xe)-phobic exothermic WIMP. }\label{Res_IV_Ex}
\end{figure}

The isospin-violating couplings~\cite{Kurylov:2003ra,Giuliani:2005my, Feng:2011vu, Cirigliano:2013zta, IVDM1} and the exothermic scatterings~\cite{Batell:2009vb,Graham:2010ca} have been the two most popular methods to ameliorate the tensions in the direct detection experiments, while the combination of these two effects has been proposed to make the CDMS-Si signal~\cite{CDMS-Si} consistent with the LUX~\cite{LUX2013} and SuperCDMS~\cite{SuperCDMS_Ge} constraints. Typical models in this class contain exothermic Ge-phobic~\cite{Gelmini:2014psa} and Xe-phobic~\cite{Chen:2014tka} WIMPs, which will be studied in detail in the present subsection. Fig.~\ref{Res_IV_Ex} shows the results for the two models with the typical parameter choices, which incorporate the 2015 data from LUX and CDMSlite experiments. Note that we only consider the cases with the mass splitting between two WIMP states to be $\delta = -200$~keV. It is well-known~\cite{Fox:2013pia, Frandsen:2014ima, Gelmini:2014psa} that the further enlargement of the mass splitting would worsen the fitting since not all the three observed events could be contained in the corresponding NR energy range. 

The rationale behind the Ge-phobic exothermic DM is that the exothermic WIMP with a mass gap $\delta = -200$~keV could push the LUX upper limit to the right of the CDMS-Si signal region~\cite{Fox:2013pia, Frandsen:2014ima, Gelmini:2014psa}, while the choice of the Ge-phobic isospin-violation parameter $\xi = -0.8$ maximally reduces the sensitivity of the germanium detector in the SuperCDMS experiments~\cite{SuperCDMS_Ge}. Therefore, it seems possible to explain all the datasets simultaneously. However, as is evident from the left plot of Fig.~\ref{Res_IV_Ex}, this mechanism cannot work any longer since the upper limits derived from the LUX2015 data exclude the whole CDMS-Si $90\%$ C.L. signal region. Note that the further enlargement of the mass gap to $\delta = -500$~keV cannot improve the conclusion, and the results are similar to those for $\delta = -200$~keV.   

In contrast, the situation for the Xe-phobic exothermic WIMP model is different, as shown in the right plot of Fig.~\ref{Res_IV_Ex}. The choice of $\xi = -0.7$ greatly lowers the potential of xenon nuclei in the direct detection of spin-independent WIMP interactions, so that SuperCDMS dominates the constraint at low WIMP mass regions relevant to the CDMS-Si signals. As a result, the current constraints for this model from null experiments remain the same as the those~\cite{Chen:2014tka,Gelmini:2014psa} 
before the LUX2015 data release, with nearly half of the CDMS-Si $68\%$ C.L. region of interest still alive. The DM mass is predicted to be around $1 \sim 2$ GeV, and the NR cross section is of ${\cal O}(10^{-42})~{\rm cm}^2$.

\subsection{Xe-phobic DM with a Light Mediator}
It has been pointed out in Ref.~\cite{Li:2014vza} that a light mediator would enhance the energy spectrum of the WIMP-nucleus recoil rate at the low-energy region favored by the CDMS-Si data, and suppress the sensitivity for high-energy NR events strongly constrained by the SuperCDMS and LUX2013 data. Together with the Xe-phobic interactions for $\xi = -0.7$, the model could partially relax the tension of the CDMS-Si, SuperCDMS and LUX2013 data. Note that the enhancement would saturate maximally when the light mediator mass becomes below the typical energy thresholds of ${\cal O}(10)$~MeV, while it would be effectively turned off for the mediator mass larger than $200$~MeV. Therefore, we only consider the case with $M_\phi = 1$~MeV in the present subsection.
\begin{figure}
\includegraphics[scale = 0.32, angle=270]{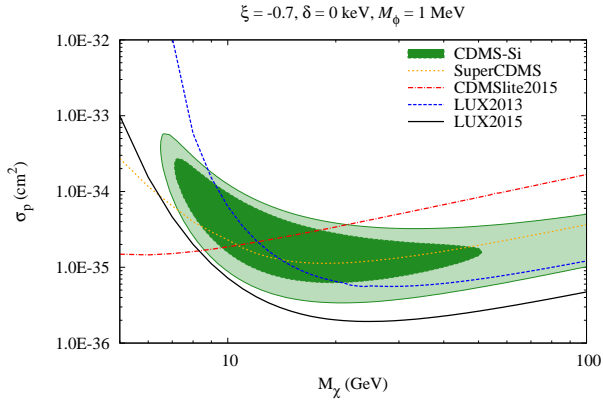}
\caption{Legend is the same as Fig.~\ref{Res_IV_Ex} but for Xe-phobic WIMP with a light mediator with its mass $m_\phi = 1$~MeV.}\label{Res_IV_LM}
\end{figure}
However, as shown in Fig.~\ref{Res_IV_LM}, such a mechanism cannot work under the constraint from the LUX2015 data, which excludes the whole CDMS-Si $90\%$ region of interest. This result is understandable in that, even though the Xe-phobic isospin-violating parameter $\xi = -0.7$ significantly weakens the LUX WIMP search ability, the xenon-based detector does not lose its sensitivity totally, due to the distribution of isotopes in the detector material. The recent upgrade of the low-energy threshold by LUX further pushes this residue sensitivity even higher.

\subsection{Exothermic DM with a Light Mediator }
\begin{figure}[ht]
\includegraphics[scale = 0.32, angle=270]{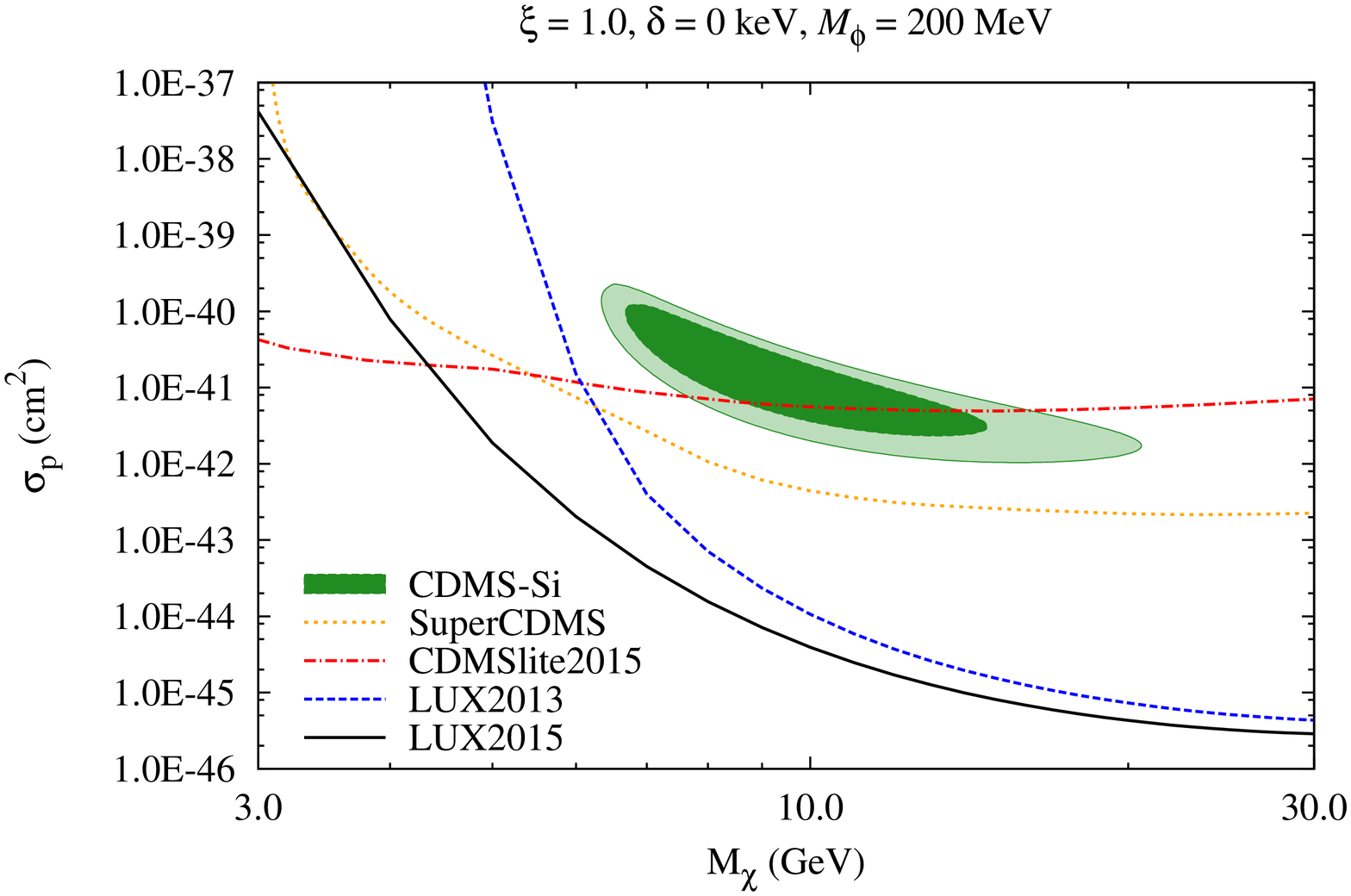}
\includegraphics[scale=0.32,angle=270]{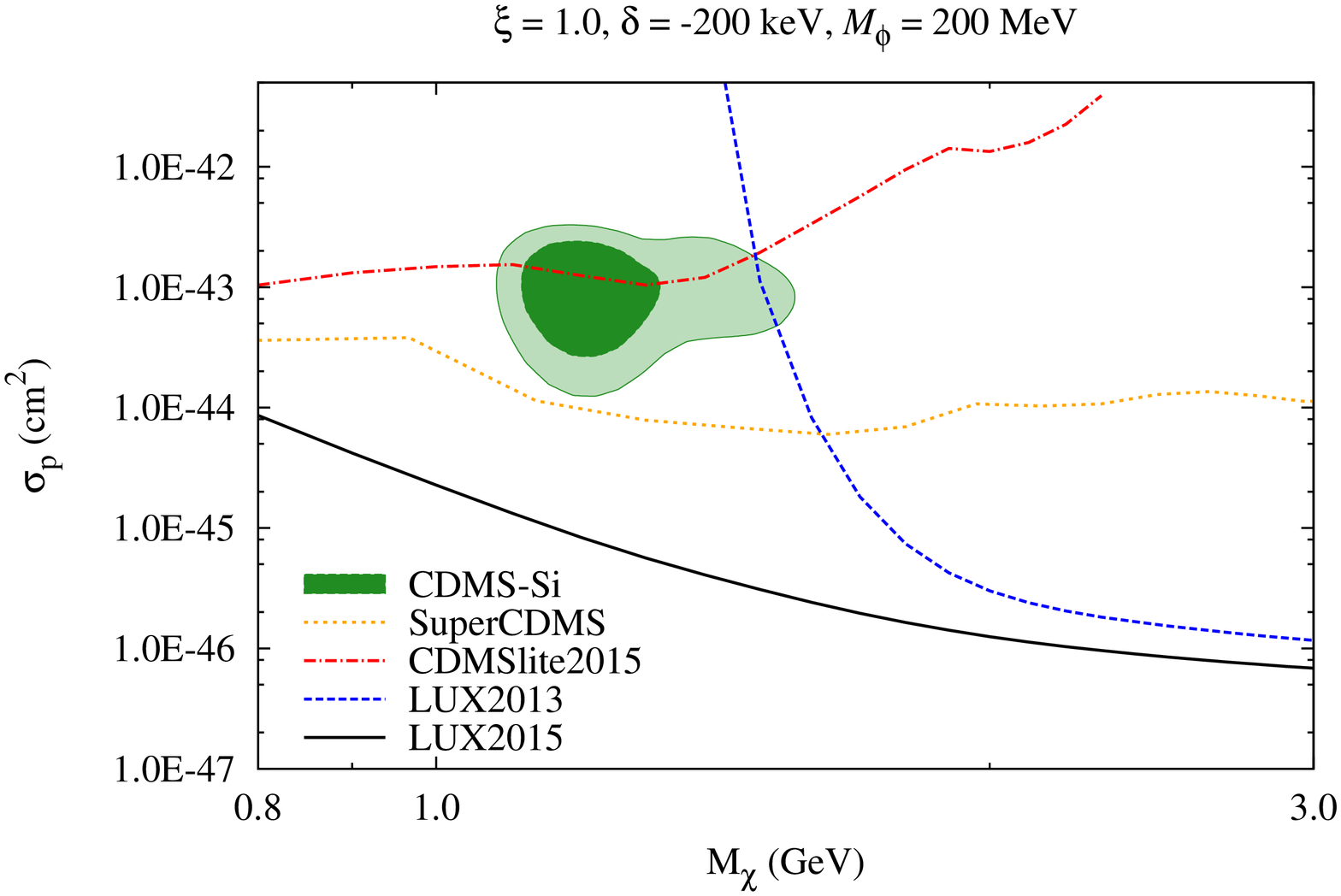}
\includegraphics[scale = 0.32, angle=270]{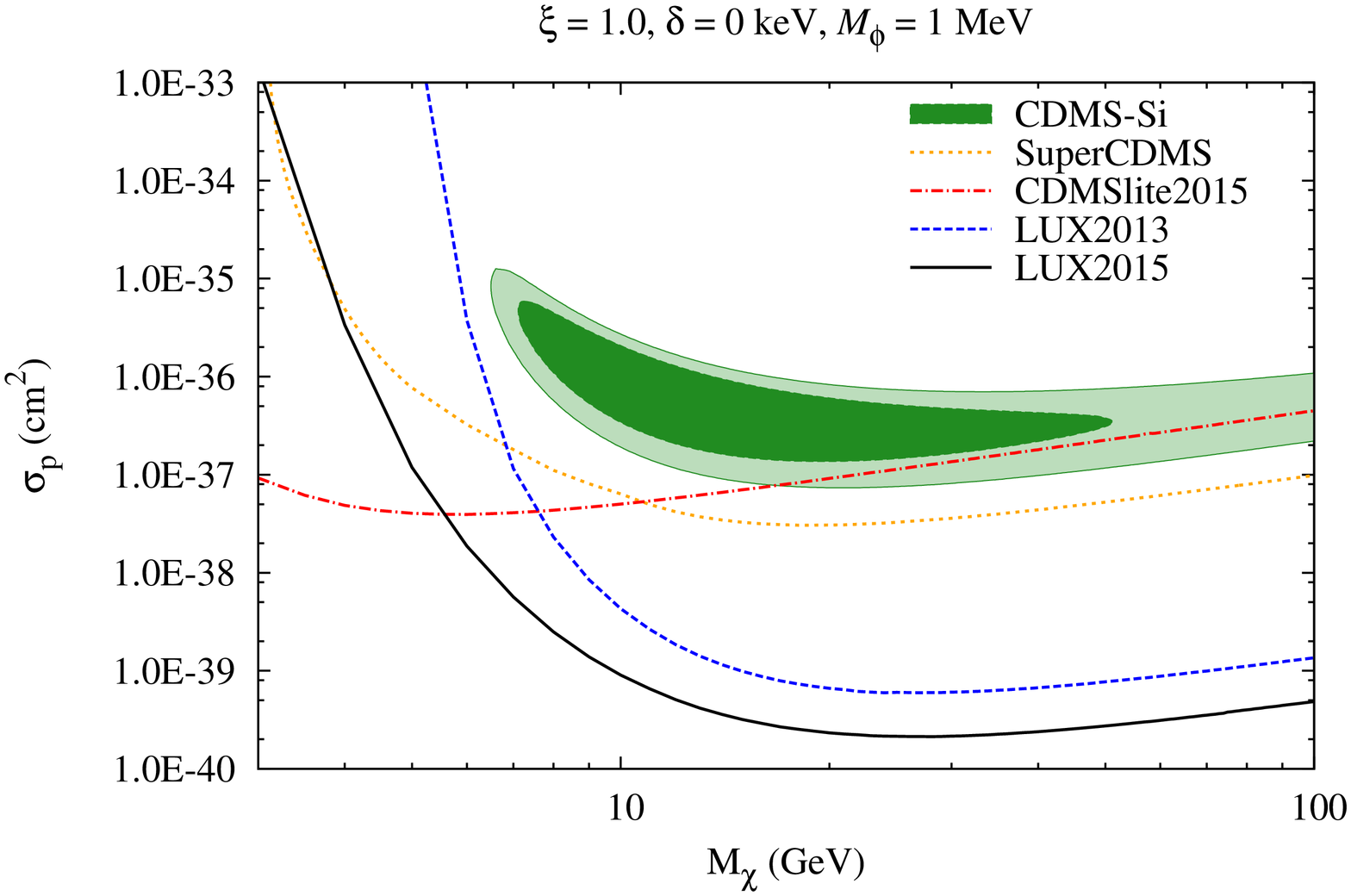}
\includegraphics[scale=0.32,angle=270]{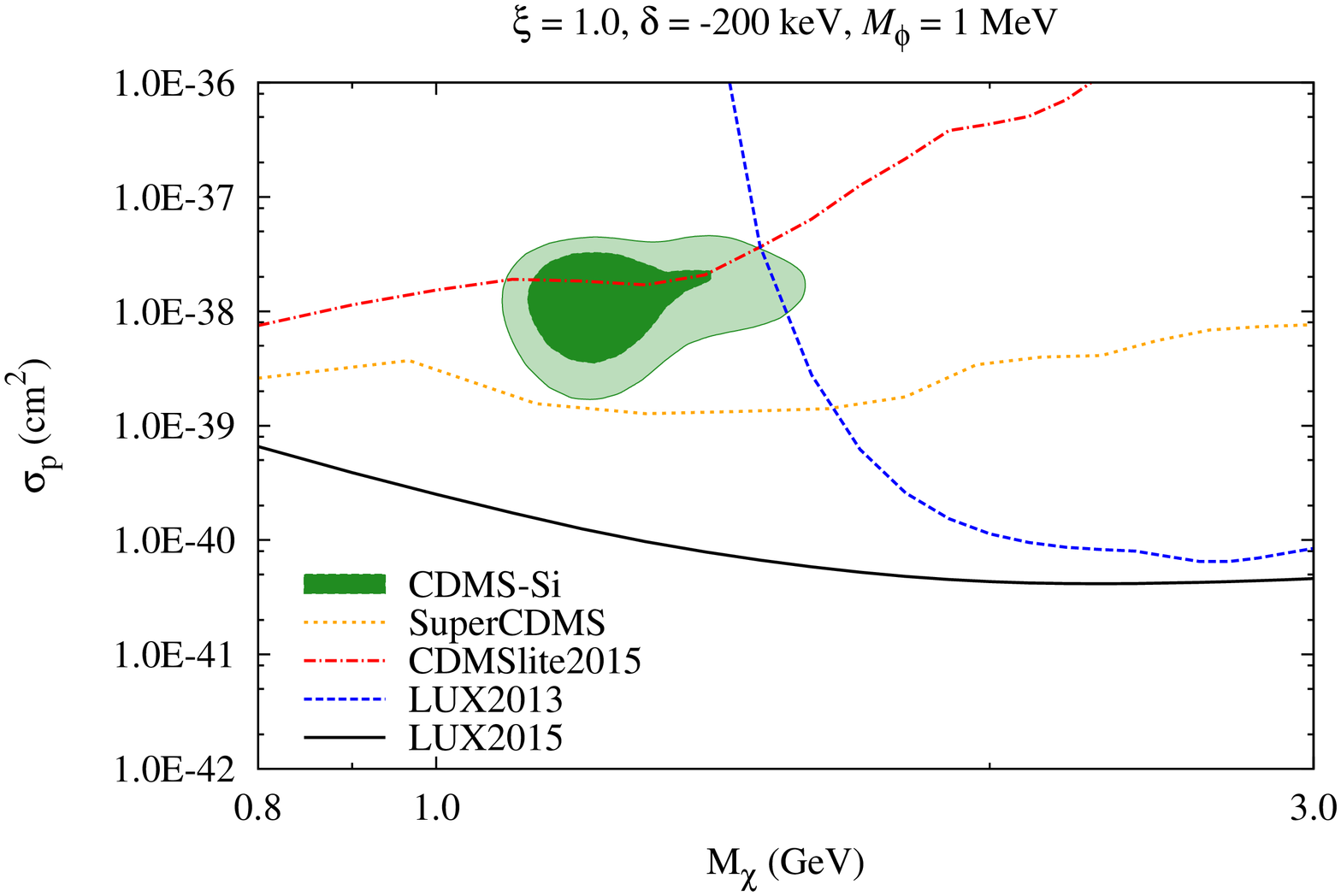}
\caption{Legend is the same as Fig.~\ref{Res_IV_Ex} but for the isospin-preserving WIMP with a mass gap $\delta = -200$~keV or/and a light mediator with $m_\phi = 1$~MeV.}\label{Res_Ex_LM}
\end{figure}

After considering the isospin violation with either exothermic WIMP-nucleus interactions or a light mediator, we wonder if an isospin-conserving exothermic WIMP model with a light mediator could also make the results in various direct detection experiments consistent with each other. Such a model has not been considered in the literature. Our final results for this model are given in the bottom right plot of Fig.~\ref{Res_Ex_LM}. In the figure, we also show those for the conventional model (Top Left) already discussed in Fig.~\ref{ResConv}, and the models with either the WIMP down-scatterings with a gap $\delta = -200$~keV (Top Right) or a light mediator with $m_\phi = 1$~MeV (Bottom Left). Note that the choice of $m_\phi = 200$~MeV corresponds to the heavy mediator case, which effectively reproduces the results of the WIMP-nucleus contact interactions.

It is seen from Fig.~\ref{Res_Ex_LM} that, for all the models, the CDMS-Si signal region is strongly disfavored by the SuperCDMS and LUX2015 data, with the most stringent exclusion limit curve from LUX. Therefore, we conclude that the idea of the exothermic DM with NRs induced by a light mediator is not possible to reduce the tensions among experiments, no matter how we choose the mediator mass $m_\phi$ and mass gap $\delta$ in the dark sector.

\subsection{Isospin-Violating Exothermic DM with a Light Mediator}
\begin{figure}[ht]
\includegraphics[scale = 0.32, angle=270]{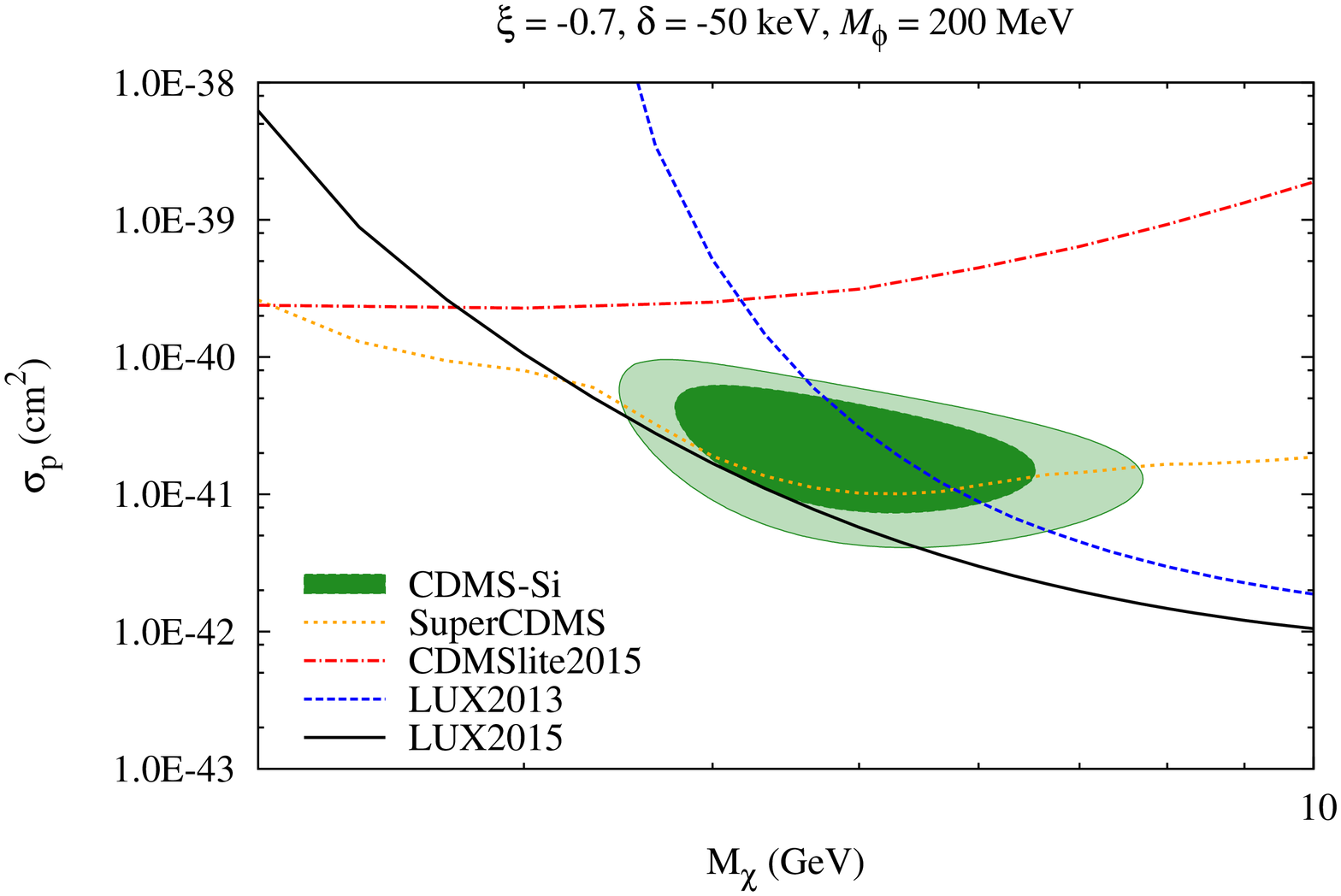}
\includegraphics[scale = 0.32, angle=270]{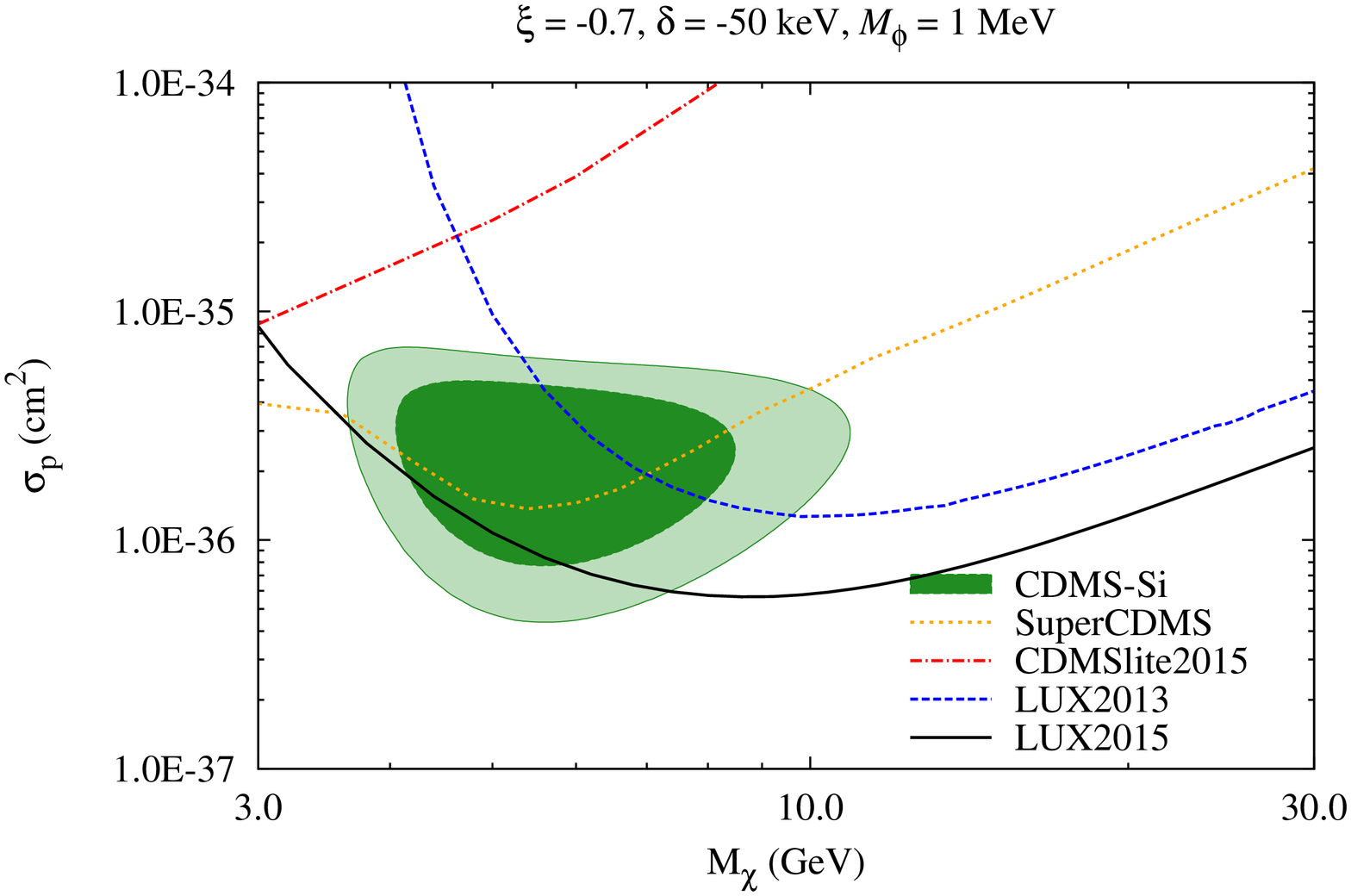}
\caption{Legend is the same as Fig.~\ref{Res_IV_Ex} but for $\xi = -0.7$, $\delta = -50$~keV and $m_\phi = 200$~MeV (Left) and $1$~MeV (Right).}\label{Res_IV_Ex50_LM}
\end{figure}

\begin{figure}[ht]
\includegraphics[scale = 0.32, angle=270]{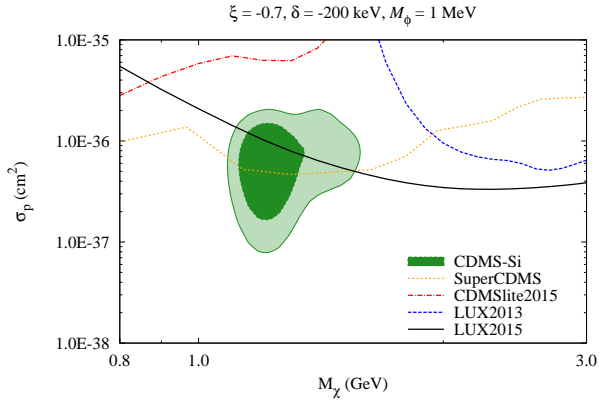}
\caption{Legend is the same as Fig.~\ref{Res_IV_Ex} but for $\xi = -0.7$, $\delta = -200$~keV and $m_\phi = 1$~MeV.}\label{Res_IV_Ex200_LM}
\end{figure}

Finally, we investigate the models which open simultaneously the aforementioned three effects: isospin-violating couplings, exothermic WIMP-nucleus scatterings, and the lightness of the mediator. We find that, compared with the isospin-violating exothermic WIMP models, the extra momentum transfer dependence due to the light mediator will generally improve the compatibility of the CDMS-Si signals with other null experiments, but the effect is only mild. Two typical examples are presented in Figs.~\ref{Res_IV_Ex50_LM} and \ref{Res_IV_Ex200_LM}.

In Fig.~\ref{Res_IV_Ex50_LM}, we explicitly show the consequences of a light mediator in the model with $\xi = -0.7$ and a relative small gap $\delta = -50$~keV. When the mediator mass is heavy with $m_\phi = 200$~MeV, in which we recover the contact WIMP-nucleus interactions, the LUX2015 constraint is very strong and excludes all of the $68\%$ C.L. CDMS-Si region of interest. The two experiments can be only compatible with each other marginally if we look at the CDMS-Si $90\%$ C.L. region. However, when the mediator becomes light, i.e., $m_\phi = 1$~MeV, a small part of the $68\%$ C.L. CDMS-Si signal region is reopened at the low WIMP mass range from 4 to 6 GeV, so that the consistency of the two experiments can be improved a little. 

Fig.~\ref{Res_IV_Ex200_LM} depicts the result when all the modification parameters are chosen to be their extremal values with $\xi =-0.7$, $\delta = 200$~keV and $m_\phi =1$~MeV. In comparison with the right plot in Fig.~\ref{Res_IV_Ex}, except for the overall boost of the WIMP-nucleon cross section from ${\cal O}(10^{-42})$ to ${\cal O}(10^{-36})~{\rm cm}^2$, the two diagrams look very similar to each other, with the best-fitting WIMP masses around $1 \sim 2$ GeV. For both cases, due to the combined effects of the Xe-phobic interaction and a large mass gap, the SuperCDMS data only cut the large cross section part of the CDMS-Si contour, where the LUX constraint is subdominant.

\begin{figure}[ht]
\includegraphics[scale = 0.6, angle=270]{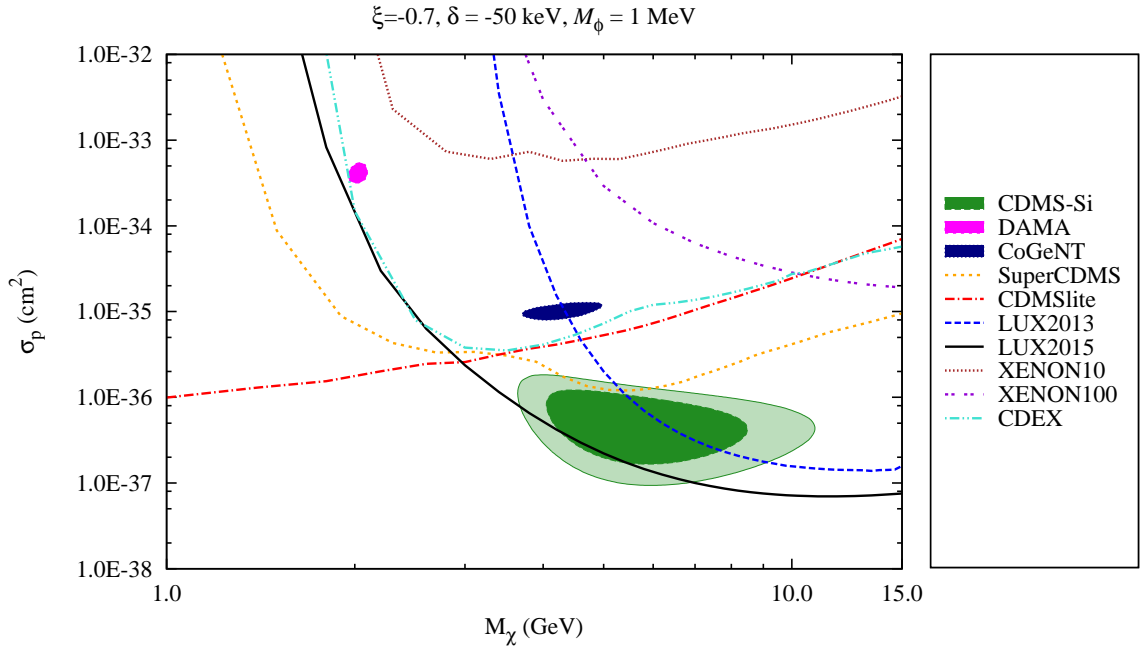}
\caption{Legend is the same as Fig.~\ref{ResConv} but for $\xi = -0.7$, $\delta = -50$~keV and $m_\phi = 1$~MeV.}\label{Res_IV_Ex50_LM_t}
\end{figure}

\begin{figure}[ht]
\includegraphics[scale = 0.6, angle=270]{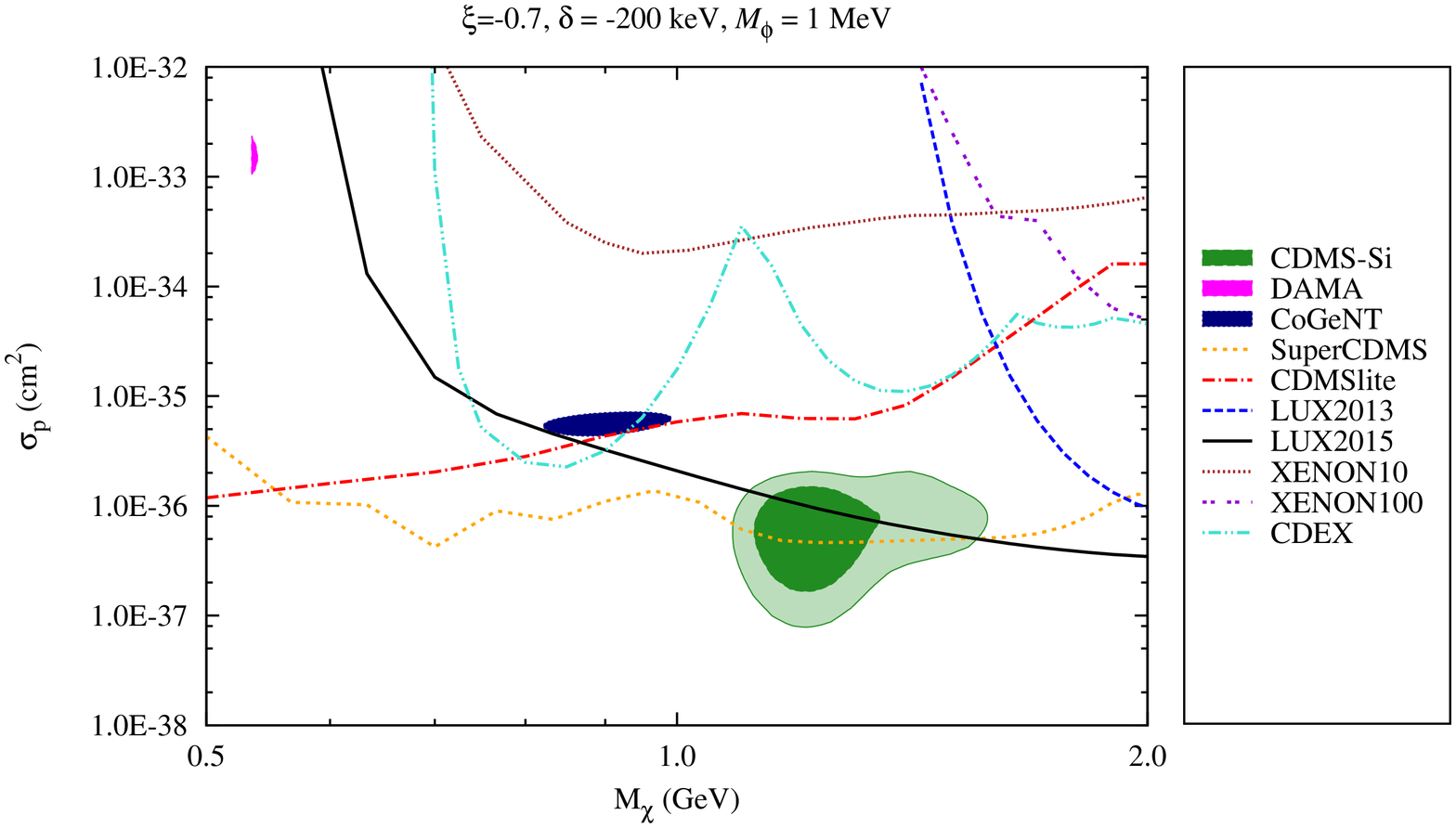}
\caption{Legend is the same as Fig.~\ref{ResConv} but for $\xi = -0.7$, $\delta = -200$~keV and $m_\phi = 1$~MeV.}
\label{Res_IV_Ex200_LM_t}
\end{figure}

In Fig.~\ref{Res_IV_Ex50_LM_t} (\ref{Res_IV_Ex200_LM_t}), we expand the right plot in Fig.~\ref{Res_IV_Ex50_LM} (\ref{Res_IV_Ex200_LM}) by including the $90\%$ C.L. regions of interest for DAMA and CoGeNT signals, as well as the $90\%$ C.L. exclusion curves for XENON10, XENON100, and CDEX-1 data. It is evident that the signal regions for DAMA and CoGeNT are still in strong conflict with the constraints from LUX, CDMSlite and SuperCDMS even for these two optimum cases.   

\section{Conclusions}\label{Sec_Conc}
We have studied the direct detections of WIMP particles with the spin-independent WIMP-nucleus interactions, and extensively examined the models which involve the isospin-violating couplings, inelastic exothermic WIMP-nucleus scatterings and/or a light mediator. These three mechanisms have been previously shown to have the potential to reduce the tension between the CDMS-Si positive signals and null experiments. We have explored the combined effects in the light of the new releases of the data from LUX and CDMSlite experiments. We have shown that it is generally difficult in finding WIMP masses and couplings consistent with all existing datasets, with the LUX2015 data usually providing the strongest constraint. In particular, we have found that the Ge-phobic exothermic DM model in Ref.~\cite{Gelmini:2014psa} and a type of Xe-phobic DM model with a light mediator in Ref.~\cite{Li:2014vza} are now ruled out up to at least $90\%$ C.L. For the models with the exothermic scatterings and the light mediator effect, we do not obtain any parameter space that is compatible with the CDMS-Si $90\%$ C.L. region of interest and satisfies the other experimental limits. The only available model is the Xe-phobic exothermic DM one with or without a light mediator. When the exothermic mass gap is relative small with $\delta = -50$~keV, the $68\%$ contour is only allowed for the light mediator case with $m_\phi = 1$~MeV, in which the WIMP mass is predicted to be in the range from 4 to 6 GeV. In comparison, when $\delta = -200$~keV, no matter the mediator is light or heavy, the WIMP mass is $1 \sim 2$ GeV, with the WIMP-nucleon cross sections of ${\cal O}(10^{-36})$ and ${\cal O}(10^{-42})~{\rm cm}^2$ for the light and heavy mediators, respectively. Nevertheless, even in this most promising case, the constraints from the SuperCDMS and LUX2015 data have already excluded a large portion of the CDMS-Si signal regions. For DAMA and CoGeNT signals, no parameter is allowed to escape the exclusion limits from LUX, SuperCDMS and CDMSlite.   

Another kind of constraints comes from the collider searches with the possible signatures as missing transverse energies plus some visible states, such as a mono-photon, mono-Z and mono-gluon. However, such investigations depend on the particular couplings between DM and SM particles and the mass of the mediator. For a heavy mediator, the limit on the spin-independent DM-nucleon cross section is typically of $\sigma_{N} \sim {\cal O} (10^{-40})~{\rm cm}^2$~\cite{collider,McCullough:2013jma}. But if the mediator is light, especially when $m_\phi < 1$~GeV, the limit is expected to be weakened significantly~\cite{collider}. 

Besides the particular effective operator in Eq.~(\ref{EOm}), which belongs to the more general Type I operators in Ref.~\cite{Li:2014vza}, there are other classes of extended spin-independent effective operators~\cite{Li:2014vza} with different momentum dependences. Furthermore, the studies in Refs.~\cite{Chang:2009yt, Fan:2010gt, Fitzpatrick:2012ix, Fitzpatrick:2012ib, Anand:2013yka, Gresham:2013mua, Catena:2016hoj} have defined more general effective operators beyond the simple spin-independent and spin-dependent models, which contain more WIMP-nucleus interactions and nuclear response functions. Clearly, it is interesting to investigate the effects of the WIMP mass gap and the light mediator for these operators too, along with the recent LUX and CDMSlite data. 

\appendix
\section{Analysis details for some relevant experiments}\label{OtherExp}
This appendix summarizes our analysis details for some relevant experiments, such as the $90\%$ C.L. signal regions from DAMA~\cite{Bernabei:2010mq} and CoGeNT~\cite{Aalseth:2012if}, as well as the XENON10~\cite{XENON10}, XENON100~\cite{XENON100} and CDEX-1~\cite{CDEX14, CDEX16} $90\%$ C.L. upper limits.
 
\subsection*{DAMA}
The DAMA/LIBRA experiments~\cite{Bernabei:2008yi,Bernabei:2010mq} have used highly radio pure NaI(Tl) scintillators as detectors aimed to search for the WIMP annual modulation signature,  detecting  the first positive signal with the C.L. as high as 8.9~$\sigma$. Following Eq.~(\ref{Modulation}), the total modulation amplitude for DAMA over the interval $[E_1, E_2]$ is given by
\begin{eqnarray}
S_{m,[E_1, E_2]} = \frac{1}{E_2 - E_1} \sum_{T = {\rm Na,~I}} c_T \int^{E_2/Q_T}_{E_1/Q_T} S_m(E_{\rm nr}) dE_{\rm nr} \,,
\end{eqnarray}
where $c_T$ represents the mass fractions of the nuclei $T =$~Na and~I and $Q_T$ is the quenching factor. In our work, we use $Q_{\rm Na} = 0.3$ and $Q_{\rm I} = 0.09$. We consider the DAMA2010 data, given in Fig.~6 of Ref.~\cite{Bernabei:2010mq}, and perform the simple minimum $\chi^2$ fitting to the data with all 36 bins, corresponding to energies from 2 to 20 keV. The resolution is taken to be 
\begin{eqnarray}
\sigma(E) = (0.048 ~{\rm keV}) \sqrt{E/{\rm keV}} + 0.0091 E\,,
\end{eqnarray} 
and the ion-channeling effects are absent in our fits. The 90$\%$ C.L. regions of interest are shown in Figs. 1, 7 and \ref{Res_IV_Ex200_LM_t}, respectively.

\subsection*{CoGeNT}
For the CoGeNT experiments, we use the data and errors given in Fig.~23 of Ref.~\cite{Aalseth:2012if} with the corrected detection efficiencies. It is evident that this dataset contains a large number of background events, so that our minimum $\chi^2$ fitting should scan over WIMP masses and WIMP-nucleon cross sections, as well as the constant background component, similar to the procedure by the Collaboration. As a result, the number of the expected events in a range $[E_1, E_2]$ is taken to be~\cite{Gresham:2013mua}
\begin{eqnarray}
N_{[E_1, E_2]} = {\rm Ex}\cdot \int^\infty_0 \frac{dR}{dE_{\rm nr}} {\Phi}(E_1, E_2, E_{\rm nr}) dE_{\rm nr} + b_{[E_1, E_2]},
\end{eqnarray}
where $b_{[E_1, E_2]}$ is the flat background, and ${\Phi}(E_1,E_2, E_{\rm nr} )$ is the factor given in Eq.~(\ref{resolution}) with $s(E_{\rm nr}) = E_{\rm nr}$, which takes into account the resolution of the experiment. For the CoGeNT data~\cite{Aalseth:2012if} below 10~keV, the energy resolution $\sigma$ can be determined by 
\begin{eqnarray}
\sigma^2 = \sigma_n^2 + 2.35^2 E_{\rm nr} \eta F\,,
\end{eqnarray}   
where $\sigma_n = 69.4$~eV, $\eta = 2.96$~eV, $F= 0.29$, and $E_{\rm nr}$ is the recoil energy in the unit of eV. 

\subsection*{XENON10}
XENON10 is a liquid xenon time-projection chamber, detecting the interaction of WIMP and xenon nuclei via the primary scintillation photons $S1$ and electrons $S2$. In Ref.~\cite{XENON10}, the XENON10 Collaboration only made the use of the electron signal $S2$ to detect the WIMP-induced NRs so that the experiment was very sensitive to the low-energy WIMP-nucleus recoils with the threshold $E_{\rm nr} \sim 1$~keV. Note that at such low NR energies, the primary scintillation signal is nearly absent. Following the Collaboration, we only adopt the highlighted events presented in Fig.~2 of Ref.~\cite{XENON10} as the NR candidates, which lay in the region from 5 to 35 electrons, corresponding to the NR interval between 1.4 to 10 keV. The NR energy is reconstructed with the electron yield $Q_y(E_{\rm nr}) = n_e/E_{\rm nr}$, shown as the solid curve in Fig.~1, by assuming a sharp cutoff at $E_{\rm nr} = 1.4$~keV. We choose a flat efficiency of $\epsilon = 94\%$ over the whole energy range, and use a parametrization of the detector energy resolution $R(E_{\rm nr}) = E_{\rm nr}/\sqrt{Q_y E_{\rm nr}}$. The expected event number in the energy range $[E_1, E_2]$ is
\begin{eqnarray}
N_{[E_1, E_2]} = {\rm Ex}\cdot \int^\infty_0 \frac{dR}{dE_{\rm nr}} \epsilon {\Phi} (E_1, E_2; E_{\rm nr}) dE_{\rm nr}\,,
\end{eqnarray}
where the exposure of the analysis in Ref.~\cite{XENON10} is ${\rm Ex} = 15$~kg-days. We apply the $p_{\rm max}$ method of Yellin~\cite{Yellin:2002xd} to obtain $90\%$ C.L. exclusion curves in Figs. 1, 7 and \ref{Res_IV_Ex200_LM_t}.

\subsection*{XENON100}
With an exposure of ${\rm Ex} = 225\times 34$~kg-days, the XENON100 Collaboration has reported~\cite{XENON100} two candidate events with the NR energies of $7.1~{\rm keV_{nr}}$ (3.3 photoelectrons (phe)) and $7.8~{\rm keV_{nr}}$ (3.8 phe) in the benchmark WIMP search region $S1\in [3~{\rm phe},20~{\rm phe}]$. Following Ref.~\cite{XENON100a}, the event number in the $S1$ range $[S1_a, S1_b]$ is taken as that in Eq.~(\ref{NLUX}), where, for XENON100 experiments, the $S2$ efficiency $\epsilon_{S2}(E_{\rm nr})$ is the red dashed line in Fig.~1~\cite{XENON100} with a hard cut at 3 keV, while the other cut acceptance $\epsilon(S1)$ as the multiplication of the dotted green and blue curves in the same plot. The averaged expected signal $S1$ given the NR energy $E_{\rm nr}$, denoted by $\nu(E_{\rm nr})$ in Eq.~(\ref{NLUX}), is also parametrized as that in Eq.~(\ref{S1transf}). For the XENON100 analysis, we pick $S_{\rm ee} = 0.58$, $S_{\rm nr} = 0.95$, $L_y = 2.28 ~{\rm phe/keV_{ee}}$, and ${\cal L}_{\rm eff}(E_{\rm nr})$ as an interpolation of the solid curve in Fig.~1 of~Ref.~\cite{PLRtest}, while $\sigma_{\rm PMT} = 0.5$~phe is the resolution of a single PMT used in the experiments. We set the 90$\%$ C.L. exclusion curves in Figs. 1, 7 and \ref{Res_IV_Ex200_LM_t} with the maximum gap statistic test method~\cite{Yellin:2002xd}.

\subsection*{CDEX-1}
The CDEX-1 experiment has a $p$-type point contact germanium detector, similar to the CoGeNT experiment, which can directly probe and constrain the CoGeNT signal region without any ambiguities. Our analysis is based on the 53.9 kg-day data presented in Fig.~3b of Ref.~\cite{CDEX14}. We assume a perfect efficiency and a perfect resolution due to the excellent performance of the detector. A flat background is also yielded from Fig.~3b of Ref.~\cite{CDEX14}, and the quenching factor for germanium NRs is taken from Ref.~\cite{Lin:2007ka}. The $90\%$ exclusion limits are obtained by performing the binned Poisson method~\cite{Green:2001xy} with bins of 0.1~keV$_{ee}$ in 
Figs.~1, 7 and \ref{Res_IV_Ex200_LM_t}. Recently, the CDEX Collaboration has updated the WIMP search data and the exclusion limits for the WIMP-nucleon spin-independent and spin-dependent cross sections with a larger exposure of 335.6 kg-days~\cite{CDEX16}. However, the improvement of the spin-independent limit is mild, especially at the low-mass region which is the main focus of the present paper. 
Clearly,  the final results do not change much due to this new data release.

\section*{Acknowledgments}
The work was supported in part by National Center for Theoretical Sciences, National Science Council (NSC-101-2112-M-007-006-MY3), and National Tsing Hua
University (104N2724E1).



\begin{thebibliography}{0}

\bibitem{PDG} 
  K.~A.~Olive {\it et al.} [Particle Data Group Collaboration],
  Chin.\ Phys.\ C {\bf 38}, 090001 (2014).
  doi:10.1088/1674-1137/38/9/090001
  
\bibitem{Planck} 
  P.~A.~R.~Ade {\it et al.} [Planck Collaboration],
  arXiv:1502.01589 [astro-ph.CO].
  
\bibitem{Steigman:1984ac} 
  G.~Steigman and M.~S.~Turner,
  Nucl.\ Phys.\ B {\bf 253}, 375 (1985).
  doi:10.1016/0550-3213(85)90537-1

\bibitem{Jungman:1995df} 
  G.~Jungman, M.~Kamionkowski and K.~Griest,
  Phys.\ Rept.\  {\bf 267}, 195 (1996)
  doi:10.1016/0370-1573(95)00058-5
  [hep-ph/9506380].
  
\bibitem{Feng:2010gw} 
  J.~L.~Feng,
  Ann.\ Rev.\ Astron.\ Astrophys.\  {\bf 48}, 495 (2010)
  doi:10.1146/annurev-astro-082708-101659
  [arXiv:1003.0904 [astro-ph.CO]].
  


\bibitem{Goodman:1984dc} 
  M.~W.~Goodman and E.~Witten,
  Phys.\ Rev.\ D {\bf 31}, 3059 (1985).
  doi:10.1103/PhysRevD.31.3059
  
\bibitem{FormFactor} 
  P.~F.~Smith and J.~D.~Lewin,
  Phys.\ Rept.\  {\bf 187}, 203 (1990).
  doi:10.1016/0370-1573(90)90081-C
  J.~D.~Lewin and P.~F.~Smith,
  Astropart.\ Phys.\  {\bf 6}, 87 (1996).
  doi:10.1016/S0927-6505(96)00047-3

\bibitem{Bernabei:2008yi} 
  R.~Bernabei {\it et al.} [DAMA Collaboration],
  Eur.\ Phys.\ J.\ C {\bf 56}, 333 (2008)
  doi:10.1140/epjc/s10052-008-0662-y
  [arXiv:0804.2741 [astro-ph]].

\bibitem{Bernabei:2010mq} 
  R.~Bernabei {\it et al.} [DAMA and LIBRA Collaborations],
  Eur.\ Phys.\ J.\ C {\bf 67}, 39 (2010)
  doi:10.1140/epjc/s10052-010-1303-9
  [arXiv:1002.1028 [astro-ph.GA]].
  
  
\bibitem{Aalseth:2010vx} 
  C.~E.~Aalseth {\it et al.} [CoGeNT Collaboration],
  Phys.\ Rev.\ Lett.\  {\bf 106}, 131301 (2011)
  doi:10.1103/PhysRevLett.106.131301
  [arXiv:1002.4703 [astro-ph.CO]].
  
\bibitem{Aalseth:2011wp} 
  C.~E.~Aalseth {\it et al.},
  Phys.\ Rev.\ Lett.\  {\bf 107}, 141301 (2011)
  doi:10.1103/PhysRevLett.107.141301
  [arXiv:1106.0650 [astro-ph.CO]].
  
\bibitem{Aalseth:2012if} 
  C.~E.~Aalseth {\it et al.} [CoGeNT Collaboration],
  Phys.\ Rev.\ D {\bf 88}, 012002 (2013)
  doi:10.1103/PhysRevD.88.012002
  [arXiv:1208.5737 [astro-ph.CO]].

\bibitem{Aalseth:2014eft} 
  C.~E.~Aalseth {\it et al.} [CoGeNT Collaboration],
  arXiv:1401.3295 [astro-ph.CO].
  
\bibitem{Aalseth:2014jpa} 
  C.~E.~Aalseth {\it et al.},
  arXiv:1401.6234 [astro-ph.CO].
  
\bibitem{CRESST-II-S} 
  G.~Angloher {\it et al.},
  Eur.\ Phys.\ J.\ C {\bf 72}, 1971 (2012)
  doi:10.1140/epjc/s10052-012-1971-8
  [arXiv:1109.0702 [astro-ph.CO]].
  
\bibitem{CDMS-Si} 
  R.~Agnese {\it et al.} [CDMS Collaboration],
  Phys.\ Rev.\ Lett.\  {\bf 111}, no. 25, 251301 (2013)
  doi:10.1103/PhysRevLett.111.251301
  [arXiv:1304.4279 [hep-ex]].
  
\bibitem{LUX2013} 
  D.~S.~Akerib {\it et al.} [LUX Collaboration],
  Phys.\ Rev.\ Lett.\  {\bf 112}, 091303 (2014)
  doi:10.1103/PhysRevLett.112.091303
  [arXiv:1310.8214 [astro-ph.CO]].
  
\bibitem{LUX2015} 
  D.~S.~Akerib {\it et al.} [LUX Collaboration],
  arXiv:1512.03506 [astro-ph.CO].



\bibitem{SuperCDMS_Ge} 
  R.~Agnese {\it et al.} [SuperCDMS Collaboration],
  Phys.\ Rev.\ Lett.\  {\bf 112}, no. 24, 241302 (2014)
  doi:10.1103/PhysRevLett.112.241302
  [arXiv:1402.7137 [hep-ex]].
  
\bibitem{Agnese:2013jaa} 
  R.~Agnese {\it et al.} [SuperCDMS Collaboration],
  Phys.\ Rev.\ Lett.\  {\bf 112}, no. 4, 041302 (2014)
  doi:10.1103/PhysRevLett.112.041302
  [arXiv:1309.3259 [physics.ins-det]].
  
\bibitem{Agnese:2015nto} 
  R.~Agnese {\it et al.} [SuperCDMS Collaboration],
  Phys.\ Rev.\ Lett.\  {\bf 116}, no. 7, 071301 (2016)
  doi:10.1103/PhysRevLett.116.071301
  [arXiv:1509.02448 [astro-ph.CO]].
  
\bibitem{XENON10} 
  J.~Angle {\it et al.} [XENON10 Collaboration],
  Phys.\ Rev.\ Lett.\  {\bf 107}, 051301 (2011)
  Erratum: [Phys.\ Rev.\ Lett.\  {\bf 110}, 249901 (2013)]
  doi:10.1103/PhysRevLett.110.249901, 10.1103/PhysRevLett.107.051301
  [arXiv:1104.3088 [astro-ph.CO]].
  
\bibitem{XENON100} 
  E.~Aprile {\it et al.} [XENON100 Collaboration],
  Phys.\ Rev.\ Lett.\  {\bf 109}, 181301 (2012)
  doi:10.1103/PhysRevLett.109.181301
  [arXiv:1207.5988 [astro-ph.CO]].

\bibitem{XENON100a} 
  E.~Aprile {\it et al.} [XENON100 Collaboration],
  Astropart.\ Phys.\  {\bf 54}, 11 (2014)
  doi:10.1016/j.astropartphys.2013.10.002
  [arXiv:1207.3458 [astro-ph.IM]].

\bibitem{CDEX14} 
  Q.~Yue {\it et al.} [CDEX Collaboration],
  Phys.\ Rev.\ D {\bf 90}, 091701 (2014)
  doi:10.1103/PhysRevD.90.091701
  [arXiv:1404.4946 [hep-ex]].
  
\bibitem{CDEX16} 
W.~Zhao {\it et al.} [CDEX Collaboration],
  Phys.\ Rev.\ D {\bf 93}, no. 9, 092003 (2016)
  doi:10.1103/PhysRevD.93.092003
  [arXiv:1601.04581 [hep-ex]].
  
\bibitem{PandaX14} 
  M.~Xiao {\it et al.} [PandaX Collaboration],
  Sci.\ China Phys.\ Mech.\ Astron.\  {\bf 57}, 2024 (2014)
  doi:10.1007/s11433-014-5598-7
  [arXiv:1408.5114 [hep-ex]].
  
\bibitem{PandaX16} 
  A.~Tan {\it et al.} [PandaX Collaboration],
  arXiv:1602.06563 [hep-ex].
  
\bibitem{CRESST-II} 
  G.~Angloher {\it et al.} [CRESST Collaboration],
  Eur.\ Phys.\ J.\ C {\bf 76}, no. 1, 25 (2016)
  doi:10.1140/epjc/s10052-016-3877-3
  [arXiv:1509.01515 [astro-ph.CO]].
  
  

\bibitem{Kurylov:2003ra} 
  A.~Kurylov and M.~Kamionkowski,
  Phys.\ Rev.\ D {\bf 69}, 063503 (2004)
  doi:10.1103/PhysRevD.69.063503
  [hep-ph/0307185].
  
\bibitem{Giuliani:2005my} 
  F.~Giuliani,
  Phys.\ Rev.\ Lett.\  {\bf 95}, 101301 (2005)
  doi:10.1103/PhysRevLett.95.101301
  [hep-ph/0504157].

\bibitem{Feng:2011vu} 
  J.~L.~Feng, J.~Kumar, D.~Marfatia and D.~Sanford,
  Phys.\ Lett.\ B {\bf 703}, 124 (2011)
  [arXiv:1102.4331 [hep-ph]].
  
\bibitem{Cirigliano:2013zta} 
  V.~Cirigliano, M.~L.~Graesser, G.~Ovanesyan and I.~M.~Shoemaker,
  Phys.\ Lett.\ B {\bf 739}, 293 (2014)
  doi:10.1016/j.physletb.2014.10.058
  [arXiv:1311.5886 [hep-ph]].

\bibitem{Savage:2008er} 
  C.~Savage, G.~Gelmini, P.~Gondolo and K.~Freese,
  JCAP {\bf 0904}, 010 (2009)
  doi:10.1088/1475-7516/2009/04/010
  [arXiv:0808.3607 [astro-ph]].
  
\bibitem{IVDM1}
  A.~L.~Fitzpatrick, D.~Hooper and K.~M.~Zurek,
  Phys.\ Rev.\ D {\bf 81}, 115005 (2010)
  doi:10.1103/PhysRevD.81.115005
  [arXiv:1003.0014 [hep-ph]].
  S.~Chang, J.~Liu, A.~Pierce, N.~Weiner and I.~Yavin,
  JCAP {\bf 1008}, 018 (2010)
  doi:10.1088/1475-7516/2010/08/018
  [arXiv:1004.0697 [hep-ph]].
  M.~T.~Frandsen, F.~Kahlhoefer, J.~March-Russell, C.~McCabe, M.~McCullough and K.~Schmidt-Hoberg,
  Phys.\ Rev.\ D {\bf 84}, 041301 (2011)
  doi:10.1103/PhysRevD.84.041301
  [arXiv:1105.3734 [hep-ph]].
  X.~G.~He, B.~Ren and J.~Tandean,
  Phys.\ Rev.\ D {\bf 85}, 093019 (2012)
  doi:10.1103/PhysRevD.85.093019, 10.1103/PhysRevD.85.119902, 10.1103/PhysRevD.85.119906
  [arXiv:1112.6364 [hep-ph]].
  N.~Okada and O.~Seto,
  Phys.\ Rev.\ D {\bf 88}, 063506 (2013)
  doi:10.1103/PhysRevD.88.063506
  [arXiv:1304.6791 [hep-ph]].
  
  
\bibitem{Frandsen:2011cg} 
  M.~T.~Frandsen, F.~Kahlhoefer, S.~Sarkar and K.~Schmidt-Hoberg,
  JHEP {\bf 1109}, 128 (2011)
  doi:10.1007/JHEP09(2011)128
  [arXiv:1107.2118 [hep-ph]].
  
\bibitem{Cline:2011zr} 
  J.~M.~Cline and A.~R.~Frey,
  Phys.\ Rev.\ D {\bf 84}, 075003 (2011)
  doi:10.1103/PhysRevD.84.075003
  [arXiv:1108.1391 [hep-ph]].
  
\bibitem{Gao:2011ka} 
  X.~Gao, Z.~Kang and T.~Li,
  JCAP {\bf 1301}, 021 (2013)
  doi:10.1088/1475-7516/2013/01/021
  [arXiv:1107.3529 [hep-ph]].

\bibitem{Chen:2014noa} 
  N.~Chen, Y.~Zhang, Q.~Wang, G.~Cacciapaglia, A.~Deandrea and L.~Panizzi,
  JHEP {\bf 1405}, 088 (2014)
  doi:10.1007/JHEP05(2014)088
  [arXiv:1403.2918 [hep-ph]].
  
  

\bibitem{Batell:2009vb} 
  B.~Batell, M.~Pospelov and A.~Ritz,
  Phys.\ Rev.\ D {\bf 79}, 115019 (2009)
  doi:10.1103/PhysRevD.79.115019
  [arXiv:0903.3396 [hep-ph]].

\bibitem{Graham:2010ca} 
  P.~W.~Graham, R.~Harnik, S.~Rajendran and P.~Saraswat,
  Phys.\ Rev.\ D {\bf 82}, 063512 (2010)
  doi:10.1103/PhysRevD.82.063512
  [arXiv:1004.0937 [hep-ph]].

\bibitem{Fox:2013pia} 
  P.~J.~Fox, G.~Jung, P.~Sorensen and N.~Weiner,
  Phys.\ Rev.\ D {\bf 89}, no. 10, 103526 (2014)
  doi:10.1103/PhysRevD.89.103526
  [arXiv:1401.0216 [hep-ph]].
  
\bibitem{McCullough:2013jma} 
  M.~McCullough and L.~Randall,
  JCAP {\bf 1310}, 058 (2013)
  doi:10.1088/1475-7516/2013/10/058
  [arXiv:1307.4095 [hep-ph]].

\bibitem{Frandsen:2014ima} 
  M.~T.~Frandsen and I.~M.~Shoemaker,
  Phys.\ Rev.\ D {\bf 89}, no. 5, 051701 (2014)
  doi:10.1103/PhysRevD.89.051701
  [arXiv:1401.0624 [hep-ph]].
  

\bibitem{Chen:2014tka} 
  N.~Chen, Q.~Wang, W.~Zhao, S.~T.~Lin, Q.~Yue and J.~Li,
  Phys.\ Lett.\ B {\bf 743}, 205 (2015)
  doi:10.1016/j.physletb.2015.02.043
  [arXiv:1404.6043 [hep-ph]].
  
\bibitem{Gelmini:2014psa} 
  G.~B.~Gelmini, A.~Georgescu and J.~H.~Huh,
  JCAP {\bf 1407}, 028 (2014)
  doi:10.1088/1475-7516/2014/07/028
  [arXiv:1404.7484 [hep-ph]].
  

  
\bibitem{Li:2014vza} 
  T.~Li, S.~Miao and Y.~F.~Zhou,
  JCAP {\bf 1503}, no. 03, 032 (2015)
  doi:10.1088/1475-7516/2015/03/032
  [arXiv:1412.6220 [hep-ph]].
  
\bibitem{Yang:2016wrl} 
  K.~C.~Yang,
  arXiv:1604.04979 [hep-ph].

\bibitem{DelNobile:2013gba} 
  E.~Del Nobile, G.~B.~Gelmini, P.~Gondolo and J.~H.~Huh,
  JCAP {\bf 1403}, 014 (2014)
  doi:10.1088/1475-7516/2014/03/014
  [arXiv:1311.4247 [hep-ph]].

\bibitem{Gresham:2013mua} 
  M.~I.~Gresham and K.~M.~Zurek,
  Phys.\ Rev.\ D {\bf 89}, no. 1, 016017 (2014)
  doi:10.1103/PhysRevD.89.016017
  [arXiv:1311.2082 [hep-ph]].
  
\bibitem{Chang:2009yt} 
  S.~Chang, A.~Pierce and N.~Weiner,
  JCAP {\bf 1001}, 006 (2010)
  doi:10.1088/1475-7516/2010/01/006
  [arXiv:0908.3192 [hep-ph]].
  
\bibitem{Fan:2010gt} 
  J.~Fan, M.~Reece and L.~T.~Wang,
  JCAP {\bf 1011}, 042 (2010)
  doi:10.1088/1475-7516/2010/11/042
  [arXiv:1008.1591 [hep-ph]].
  
\bibitem{Fitzpatrick:2012ix} 
  A.~L.~Fitzpatrick, W.~Haxton, E.~Katz, N.~Lubbers and Y.~Xu,
  JCAP {\bf 1302}, 004 (2013)
  doi:10.1088/1475-7516/2013/02/004
  [arXiv:1203.3542 [hep-ph]].
  
\bibitem{Fitzpatrick:2012ib} 
  A.~L.~Fitzpatrick, W.~Haxton, E.~Katz, N.~Lubbers and Y.~Xu,
  arXiv:1211.2818 [hep-ph].
  
\bibitem{Anand:2013yka} 
  N.~Anand, A.~L.~Fitzpatrick and W.~C.~Haxton,
  Phys.\ Rev.\ C {\bf 89}, no. 6, 065501 (2014)
  doi:10.1103/PhysRevC.89.065501
  [arXiv:1308.6288 [hep-ph]].

  
\bibitem{Catena:2016hoj} 
  R.~Catena, A.~Ibarra and S.~Wild,
  arXiv:1602.04074 [hep-ph].

\bibitem{DDDM1} 
  J.~Fan, A.~Katz, L.~Randall and M.~Reece,
  Phys.\ Rev.\ Lett.\  {\bf 110}, no. 21, 211302 (2013)
  doi:10.1103/PhysRevLett.110.211302
  [arXiv:1303.3271 [hep-ph]].
  
\bibitem{DDDM2} 
  J.~Fan, A.~Katz, L.~Randall and M.~Reece,
  Phys.\ Dark Univ.\  {\bf 2}, 139 (2013)
  doi:10.1016/j.dark.2013.07.001
  [arXiv:1303.1521 [astro-ph.CO]].


\bibitem{Freese:1987wu} 
  K.~Freese, J.~A.~Frieman and A.~Gould,
  Phys.\ Rev.\ D {\bf 37}, 3388 (1988).
  doi:10.1103/PhysRevD.37.3388
  
\bibitem{Smith:2006ym} 
  M.~C.~Smith {\it et al.},
  Mon.\ Not.\ Roy.\ Astron.\ Soc.\  {\bf 379}, 755 (2007)
  doi:10.1111/j.1365-2966.2007.11964.x
  [astro-ph/0611671].

\bibitem{Drukier:1986tm} 
  A.~K.~Drukier, K.~Freese and D.~N.~Spergel,
  Phys.\ Rev.\ D {\bf 33}, 3495 (1986).
  doi:10.1103/PhysRevD.33.3495

  
\bibitem{CDMS_Si_bkg} 
CDMS Collaboration, K.~McCarthy, ``Dark matter search results from the silicon detectors of the cryogenic dark matter search experiment,'' 
available at http://cdms.berkeley.edu/presentations/APS$\_$CDMS$\_$Si$\_$2013$\_$McCarthy.pdf.

\bibitem{Barlow:1990vc} 
  R.~J.~Barlow,
  Nucl.\ Instrum.\ Meth.\ A {\bf 297}, 496 (1990).
  doi:10.1016/0168-9002(90)91334-8
  
  
\bibitem{Yellin:2002xd} 
  S.~Yellin,
  Phys.\ Rev.\ D {\bf 66}, 032005 (2002)
  doi:10.1103/PhysRevD.66.032005
  [physics/0203002].

\bibitem{Lindhard}
 J.~Lindhard, V.~Nielsen, M.~Scharff, and P.~V.~Thomsen, K. Dan. Vidensk. Selsk. Mat. Fys. Medd. {\bf 33}, 10 (1963)
 available at http://www.sdu.dk/Bibliotek/matfys.  
  
\bibitem{Sorensen:2011bd} 
  P.~Sorensen and C.~E.~Dahl,
  Phys.\ Rev.\ D {\bf 83}, 063501 (2011)
  doi:10.1103/PhysRevD.83.063501
  [arXiv:1101.6080 [astro-ph.IM]].

\bibitem{PLRtest} 
  E.~Aprile {\it et al.} [XENON100 Collaboration],
  Phys.\ Rev.\ D {\bf 84}, 052003 (2011)
  doi:10.1103/PhysRevD.84.052003
  [arXiv:1103.0303 [hep-ex]].
  
\bibitem{LUXLightYield} LUX Collaboration, D.~McKinsey and R.~Gaitskell, First Science Result from the LUX Dark Matter Experiment, http://luxdarkmatter.org, Oct.~30, 2013.

\bibitem{collider} Primulando, FERMILAB-THESIS-2012-31.

\bibitem{Lin:2007ka} 
  S.~T.~Lin {\it et al.} [TEXONO Collaboration],
  Phys.\ Rev.\ D {\bf 79}, 061101 (2009)
  doi:10.1103/PhysRevD.79.061101
  [arXiv:0712.1645 [hep-ex]].

\bibitem{Green:2001xy} 
  A.~M.~Green,
  Phys.\ Rev.\ D {\bf 65}, 023520 (2002)
  [astro-ph/0106555].




\end{thebibliography}
\end{document}